\documentclass[12pt]{article}
\usepackage[OT1]{fontenc}
\usepackage[utf8]{inputenc}
\usepackage{amsmath,amsthm,amssymb}
\usepackage{graphicx,psfrag,epsf}
\usepackage{color}
\usepackage[pdftex]{hyperref}
\usepackage{url}
\usepackage{natbib}
\bibpunct{(}{)}{,}{a}{}{;}

\usepackage{multirow}
\usepackage{color,colortbl}
\definecolor{lightgray}{rgb}{.9,.9,1}

\newcommand{\blind}{0}

\newcommand{\bfy}{\mathbf{y}}

\newcommand{\bfX}{\mathbf{X}}

\newcommand{\bv}{\mathbf{v}}

\newcommand{\bx}{\mathbf{x}}

\newcommand{\bX}{\mathbf{X}}

\newcommand{\by}{\mathbf{y}}

\newcommand{\bfbeta}{\boldsymbol{\beta}}

\newcommand{\bftau}{\boldsymbol{\tau}}

\newcommand{\bfLambda}{\boldsymbol{\Lambda}}

\newcommand{\bfSigma}{\boldsymbol{\Sigma}}

\newcommand{\hatbeta}{\hat{\beta}}
\newcommand{\hatbfbeta}{\,\hat{\!\bfbeta}}

\makeatletter
\newcommand*{\rom}[1]{\expandafter\@slowromancap\romannumeral #1@}
\makeatother

\DeclareMathOperator*{\argmin}{arg\,min}
\newcommand{\norm}[2][2]{\left\|#2\right\|_{#1}}

\DeclareMathOperator{\SNR}{SNR}
\DeclareMathOperator{\FDR}{FDR}
\DeclareMathOperator{\SEN}{SEN}

\usepackage{fge}
\newcommand{\D}{\mathcal{D}}

\newcommand{\Supp}{\mathcal{S}}
\newcommand{\Sall}{{\lbrace \Supp_{\lambda} \rbrace}_{\lambda \in \Lambda}}
\newcommand{\Fall}{{\lbrace \mathcal{F}_{\lambda} \rbrace}_{\lambda \in \Lambda}}
\newcommand{\Sest}{\Supp_{\hat{\lambda}}}

\newcommand{\Shat}{\hat{\Supp}}
\newcommand{\Sreal}{\Supp^{\star}}

\newcommand{\betajreal}{\beta_{j}^\star}

\newcommand{\hatyhyponepermute}{\hat{\by}_{1}^{(b)}}

\newcommand{\Xperm}{{\bfX^{(b)}}}
\newcommand{\vperm}{\bv^{(b)}}
\newcommand{\aperm}{a^{(b)}}
\newcommand{\betaperm}{\hat{\bfbeta}^{(b)}}
\newcommand{\betanotj}{\hat{\bfbeta}_{\fgebackslash j}}
\newcommand{\varnotj}{\bfX_{\fgebackslash j}}
\newcommand{\varj}{\bx_{j}^{(b)}}
\newcommand{\lambdanotj}{{\bfLambda_{\fgebackslash j}}}

\newif\ifold\oldfalse

\begin{document}  

\def\spacingset#1{\renewcommand{\baselinestretch}%
{#1}\small\normalsize} \spacingset{1}
\if0\blind
{
  \title{\bf Beyond Support \\ in Two-Stage Variable Selection}
  \author{Jean-Michel~B\'ecu\thanks{Corresponding author.} \thanks{J.-M. B\'ecu and Y. Grandvalet are with 
  Sorbonne universit\'es, 
  Universit\'e de technologie de Compi\`egne, CNRS, Heudiasyc
  UMR 7253, CS 60 319, 60 203 Compi\`egne cedex, France,
  e-mail: jean-michel.becu@hds.utc.fr, yves.grandvalet@utc.fr.} ,        
        Yves~Grandvalet\footnotemark[2] ,\\
        Christophe~Ambroise\thanks{
  C. Ambroise and C. Dalmasso are with the LaMME, Universit\'e d'\'Evry
  val d'Essonne, 91000 \'Evry, France, email : christophe.ambroise@genopole.cnrs.fr, cyril.dalmasso@genopole.cnrs.fr.}~ and
        Cyril~Dalmasso\footnotemark[3] .}
  \maketitle
} \fi

\if1\blind
{
  \bigskip
  \bigskip
  \bigskip
  \begin{center}
    {\LARGE\bf Beyond Support \\ in Two-Stage Variable Selection}
\end{center}
  \medskip
} \fi


\bigskip
\begin{abstract}
  Numerous variable selection methods rely on a two-stage procedure, where a
  sparsity-inducing penalty is used in the first stage to predict the support,
  which is then conveyed to the second stage for estimation or inference
  purposes.  In this framework, the first stage screens variables to find a set
  of possibly relevant variables and the second stage operates on this set of
  candidate variables, to improve estimation accuracy or to assess the
  uncertainty associated to the selection of variables.
  We advocate that more information can be conveyed from the first stage to the
  second one: we use the magnitude of the coefficients estimated in the first
  stage to define an adaptive penalty that is applied at the second stage.
  We give two examples of procedures that can benefit from the proposed transfer
  of information, in estimation and inference problems respectively. 
  Extensive simulations demonstrate that this transfer is particularly efficient
  when  each stage operates on distinct subsamples.
  This separation plays a crucial role for the computation of calibrated
  $p$-values, allowing to control the False Discovery Rate.
  In this setup, the proposed transfer results in sensitivity gains ranging 
  from 50\% to 100\% compared to state-of-the-art.
\end{abstract}

\noindent%
{\it Keywords:}  Linear model, Lasso, Variable selection, $p$-values, False discovery rate, Screen and clean
\vfill

\newpage
\spacingset{1.45} 
\section{Introduction}

The selection of explanatory variables has attracted much attention these last
two decades, particularly for high-dimensional data, where the number of
variables is greater than the number of observations.
This type of problem arises in a variety of
domains,
including image analysis \citep{Wang2008}, 
chemometry \citep{Chong05} and genomics \citep{Xing01,Ambroise02,Anders10}. 
Since the development of the sparse estimators derived from $\ell_{1}$ 
penalties such as the Lasso \citep{Tibshirani96} or the Dantzig selector
\citep{Candes07}, sparse models have been shown to be able to recover the subset 
of relevant variables in various situations \citep[see,
e.g.][]{Candes07,Verzelen12,Buhlman13, Tenenhaus14}.

However, the conditions for support recovery are quite stringent and difficult
to assess in practice. 
Furthermore, the strength of the penalty to be applied differs between the
problem of model selection, targeting the recovery of the support of regression
coefficients, and the problem of estimation, targeting the accuracy of these
coefficients.
As a result, numerous variable selection methods rely on a two-stage
procedure, where the Lasso is used in the first stage to predict the support, 
which is then conveyed to the second stage for estimation or inference purposes.
In this framework, the first stage screens variables to find a set of possibly
relevant variables and the second stage operates on this set of candidate
variables, to improve estimation accuracy or to assess the uncertainty
associated to the selection of variables.

This strategy has been proposed to correct for the estimation bias of the
Lasso coefficients, with several variants in the second stage.
The latter may then be performed by ordinary least squares (OLS) regression for
the LARS/OLS Hybrid of \cite{Efron04} \citep[see also][]{Belloni13}, by the Lasso for the Relaxed Lasso of
\cite{Meinshausen07}, by modified least squares or ridge regression for
\cite{Liu14}, or with ``any reasonable regression method'' for the marginal
bridge of \cite{Huang08}.

The same strategy has been proposed to perform variable selection with
statistical guarantees by \cite{Wasserman09}, whose approach was pursued by
\cite{Meinshausen09}.  The first stage performs variable selection by Lasso or
other regression methods on a subset of data.  It is followed by a second stage
relying on the OLS, on the remaining subset of data, to test the relevance of
these selected variables.
\footnote{%
  In their two-stage procedure, \cite{Liu14} also proposed to construct
  confidence regions and to conduct hypothesis testing by bootstrapping
  residuals.  
  Their approach fundamentally differs from \cite{Wasserman09}, in that
  inference does not rely on the two-stage procedure itself, but on the
  properties of the estimator obtained in the second stage.
}

To summarize, the first stage of these approaches screens variables and transfers
the estimated support of variables to the second stage for a more focused
in-depth analysis.
In this paper, we advocate that more information can be conveyed from the first 
stage to the second one, by using the magnitude of the coefficients estimated in 
the first stage.  
Improving this information transfer is essential in the so-called the 
large $p$ small $n$ designs which are typical in genomic applications. 
The magnitude of regression coefficients is routinely interpreted as a quantitative gauge of
relevance in statistical analysis, can be used to define an adaptive penalty, 
following alternative views of sparsity-inducing penalties.
These views may originate from variational methods regarding optimization, or
from hierarchical Bayesian models, as detailed in Section~\ref{sec:adaptiveridge}.
In Sections~\ref{sec:estimation} and~\ref{sec:screen_clean}, we give two
examples of procedures that can benefit from the proposed transfer of magnitude
in estimation and inference problems respectively.
The actual benefits are empirically demonstrated in Section~\ref{sec:numerical}.

\section{Beyond Support: Magnitude}
\label{sec:adaptiveridge}

We consider the following high-dimensional sparse linear regression model:
\begin{equation*}
   \by = \bX \bfbeta^\star + \boldsymbol{\varepsilon}
   \enspace,
\end{equation*}
where $\by = (y_1,\cdots, y_n)^t$ is the vector of responses,
$\bX$ is the $n\times p$ design matrix with $p\gg n$, 
$\bfbeta^\star$ is the sparse $p$-dimensional vector of unknown parameters,
and $\boldsymbol{\varepsilon}$ is a $n$-dimensional vector of independent random
variables of mean zero and variance $\sigma^2$.

We discuss here two-stage approaches relying on a first screening of variables
based on the Lasso,
which is nowadays widely used to tackle simultaneously variable estimation and
selection.~\footnote{%
Though many sparsity-inducing penalties, such as the Elastic-Net, the
group-Lasso or the fused-Lasso lend themselves to the approach proposed here, we
will stick to the simple Lasso penalty throughout the paper.
}
The original Lasso estimator is defined as:
\begin{equation}
  \label{eq:lasso}
  \hatbfbeta{(\lambda)} = \argmin_{\bfbeta\in\mathbb{R}^p}
  J (\bfbeta) + \lambda \norm[1]{\bfbeta}
  \enspace, 
\end{equation}
where $\lambda$ is a hyper-parameter, and $J(\bfbeta)$ is the
data-fitting term.
Throughout this paper, we will discuss regression problems for which
$J(\bfbeta)$ is defined as
\begin{equation*}
  J (\bfbeta) = \dfrac{1}{2} \norm{\bX\bfbeta - \bfy}^2
  \enspace,
\end{equation*}
but, except for the numerical acceleration tricks mentioned in
Appendix~\ref{sec:implementation:details}, the overall feature selection process
may be applied to any other form of $J(\bfbeta)$, thus allowing to address 
classification problems.

Our approach relies on an alternative view of the Lasso, seen as an
adaptive-$\ell_2$ penalization scheme, following a viewpoint that has been 
mostly taken for optimization purposes \citep{Grandvalet98a,Grandvalet98b,Bach12}. 
It may be formalized as a variational form of the Lasso:
\begin{equation}
  \label{eq:lasso:variational}
  \begin{split}
  \min_{\bfbeta\in\mathbb{R}^p,\bftau\in\mathbb{R}^p} ~ &
  J (\bfbeta) + \lambda \sum\limits_{j=1}^p \frac{1}{\tau_{j}} \beta_{j}^{2} \\
  \text{s.~t.}~ & \sum\limits_{j=1}^p \tau_j -  \sum\limits_{j=1}^p \left|\beta_{j}\right| \leq 0 
                \enspace,\enspace \tau_j \geq 0 \enspace,\ j=1,\ldots,p 
  \enspace.
 \end{split}
\end{equation}
The variable $\bftau$ introduced in this formulation, which adapts the $\ell_2$
penalty to the data, can be shown to lead to the following adaptive-ridge 
penalty:
\begin{equation}
  \label{eq:adaptive-ridge_penalty}
  \sum\limits_{j=1}^p \frac{\lambda}{|\hatbeta_j{(\lambda)}|} \beta_{j}^{2}
  \enspace,
\end{equation}
where the coefficients $\hatbeta_j{(\lambda)}$ are the solution to 
the Lasso problem \eqref{eq:lasso}.

Using this adaptive-$\ell_2$ penalty returns the original Lasso estimator
(see proof in Appendix \ref{sec:equivalence}).
This equivalence is instrumental here for defining the data-dependent penalty 
\eqref{eq:adaptive-ridge_penalty},
implicitly determined in the first stage through the Lasso estimate, that
will also be applied in the second stage.
In this process, our primary aim is to retain the magnitude of the coefficients
of $\hatbfbeta{({\lambda})}$ in addition to the support $\Supp_{\lambda} = 
\{j\in\{1,...,p\}|\hatbeta_j{(\lambda)}\neq0\}$: the
coefficients estimated to be small in the first stage will thus also be encouraged to be
also small in the second stage, whereas the largest ones will be less penalized.

The variational form of the Lasso can be interpreted as a hierarchical 
model in the Bayesian framework~\citep{Grandvalet98b}.  
In this interpretation, together with $\lambda$ and the noise 
variance, the $\tau_j$ parameters of Problem \eqref{eq:lasso:variational} define 
the diagonal covariance matrix of a centered Gaussian prior on $\bfbeta$ 
(assuming a Gaussian noise model on $\bfy$).
Hence, ``freezing'' the $\tau_j$ parameters at the first stage of a two-stage 
approach can be interpreted as picking the parameters of the Gaussian prior on
$\bfbeta$ to be used at the second stage.

\section{A Two-Stage Estimation Procedure: Lasso+Ridge}
\label{sec:estimation}

In sparse linear regression models, several theoretical results state conditions
that ensure asymptotical support recovery, that is, the recovery of the subset
of all relevant explanatory variables.
%
One of the main result reveals a necessary and sufficient condition for the
selection property of $\ell_1$-regularized least squares.  Several variants of
this condition have been proposed, such as the irrepresentable condition, the 
restricted eigenvalue condition, or the mutual incoherence condition.
In a nutshell, this type of condition states that the subset of truly
effective variables can be retrieved exactly, provided the relevant and 
irrelevant covariates are not too strongly correlated. 
However, the rate of convergence of the Lasso may be slow and many noise
variables are selected if the estimator is chosen by a predictive criterion such
as cross-validation \citep{Meinshausen07}.
These observations motivated the proposal of several two-stage procedures
\citep{Efron04,Meinshausen07,Huang08,Belloni13,Liu14}.
They produce models with faster convergence, smaller bias, and even, under more
restrictive assumptions, oracle guarantees.


In this paper, we experimentally investigate the large $p$ small $n$ designs for
the Lasso+OLS \citep{Efron04,Belloni13} and Lasso+Ridge \citep{Liu14}
procedures, comparing them to a variant based on adaptive ridge. 
We do not work out the proofs of \cite{Liu14} to show the consistency of the 
adaptive ridge variant, since we believe that this transposition would be of low
interest.
 
\subsection{Original Lasso+OLS and Lasso+Ridge Procedures}

In these two-stage procedures, the support $\Supp_{\lambda}$ of the sparse Lasso
estimator $\hatbfbeta{(\lambda)}$ of Equation~\eqref{eq:lasso} defines the set
of possibly relevant variables.  Then, either ordinary least squares or ridge
regression is applied to the selected predictors:
\begin{equation*}
  \label{eq:ridge}
  \tilde{\bfbeta}{(\lambda,\mu)} = 
  \argmin_{\bfbeta\in\mathbb{R}^{p}:\beta_{j}=0, j\notin\Supp_{\lambda}}
  J (\bfbeta) + \mu \norm[2]{\bfbeta}^{2}
  \enspace, 
\end{equation*}
where we have the Lasso+OLS for $\mu=0$.

\cite{Belloni13} and \cite{Liu14} work out the rates that should govern the
decay of the Lasso penalty parameter $\lambda$ for Lasso+OLS and Lasso+Ridge
respectively, but they do not propose a practical means of setting the constants
so as to define the actual estimator.
In their experimental section, \cite{Liu14} however compute $\lambda$ by
cross-validation, while the ridge parameter $\mu$ is set to $1/n$, thereby
following the rate decay that theoretically enjoys good estimation and
prediction performances.

\subsection{Lasso+Adaptive Ridge Procedure}

In practice, the actual choice of the penalization parameters $\lambda$ and 
$\mu$ is very important regarding performances. 
Cross-validation is commonly used to estimate the penalty parameter $\lambda$ of 
the Lasso estimator, and we follow \cite{Liu14} in using this scheme for 
setting $\lambda$ for Lasso+OLS, Lasso+Ridge and Lasso+Adaptive Ridge, defined
as:
\begin{equation*}
  \label{eq:adpativeridge}
  \tilde{\bfbeta}{(\lambda,\mu)} = 
  \argmin_{\bfbeta\in\mathbb{R}^{p}:\beta_{j}=0, j\notin\Supp_{\lambda}}
  J (\bfbeta) + \mu \sum\limits_{j=1}^p \frac{\lambda}{|\hatbeta_j{(\lambda)}|} \beta_{j}^{2}
  \enspace,
\end{equation*}
where $\hatbeta_{j}(\lambda)$ are the regression coefficients obtained by the
Lasso with penalty parameter $\lambda$.
Then, as setting arbitrarily $\mu=1/n$ can lead to very bad performances for
Lasso+Ridge or Lasso+Adaptive Ridge, we also chose to set $\mu$ by 
cross-validation.

Note that, if applied naively, this serial selection process is prone to
overfitting, in the sense that the
variables selected by the Lasso are likely to be correlated with the response
variable, resulting in optimistic conclusions regarding variable importance,
a phenomenon known as Freedman's paradox in model selection
\citep[see][]{Freedman83b}. 
Our protocol consists in cross-validating the complete serial
process to select $\mu$ once $\lambda$ has been chosen in the screening stage of
the procedure (that is, $\lambda$ is fixed, but $\Supp_{\lambda}$ is recomputed
at each fold of the cross-validation process).
Finally, following \cite{Meinshausen07}, we set jointly 
$\lambda$ and $\mu$ by cross-validation, so that the $\lambda$ parameter of the 
Lasso screening is not optimized so as to minimize the expected prediction error
of the Lasso estimator itself, but it is optimized so as to optimize this error
for the Lasso+Adaptive Ridge estimator.

\section{A Two-Stage Inference Procedure: Screen and Clean}
\label{sec:screen_clean}

When interpretability is a key issue, it is essential to take into account the
uncertainty associated to the selection of variables inferred from limited data.
Indeed, this assessment is critical before investigating possible effects, 
since there is no way to ascertain that the support is identifiable.
Indeed, in practice, the irrepresentable condition and related conditions
cannot be checked \citep{Buhlman13}.

A classical way to assess the predictor uncertainty consists in testing the
significance of each predictor by statistical hypothesis testing and the derived
$p$-values.
Although $p$-values have a number of disadvantages and are prone to possible
misinterpretations, it is the numerical indicator that most end-users rely upon
when selecting predictors in high-dimensional context.
Well-established and routinely used selection methods in genomics are univariate
\citep{Balding2006}.
Although more powerful, multivariate approaches suffer from instability and lack
of usual measure of uncertainty.
It is only recently that  means for computing $p$-values or confidence
intervals in the high-dimensional
regression setup were proposed, originating with the work of
\citet{Wasserman09} and  followed by others \citep{Meinshausen09,Buhlman13,Liu14}.
From a practical point of view, these recent developments are essential for
convincing practitioners of the benefits of multivariate sparse regression
models \citep{Boulesteix14}.
Here, we build on the seminal work of \citet{Wasserman09}.
We propose to introduce adaptive ridge in the cleaning stage to transfer more
information from the screening stage to the cleaning stage, and thus to make a
more extensive use of the subsample of the original data reserved for screening 
purposes.

\subsection{Original Screen and Clean Procedure}
\label{sec:screen_clean:original}

The procedure considers a series of sparse models $\Fall$, indexed by a
parameter $\lambda\in\Lambda$, which
may represent a penalty parameter for regularization methods or a size
constraint for subset selection methods.
The screening stage consists of two steps.
In the first step, each model $\mathcal{F}_{\lambda}$ is fitted to (part of) the
data, thereby selecting a set of possibly relevant variables, that is, the
support of the model $\Supp_{\lambda}$.
Then, in the second step, a model selection procedure chooses a single model
$\mathcal{F}_{\hat{\lambda}}$ with its associated $\Sest$.
Next, the cleaning stage eliminates possibly irrelevant variables from
$\Sest$, resulting in the set  $\Shat$ that provably controls the type
one error rate.
The original procedure relies on three independent subsamples of the original
data $\D = \D_1 \cup \D_{2} \cup \D_{3}$, so as to ensure the consistency of the
overall process.
The following chart summarizes this procedure, showing the actual use of data
that is made at each step:
\begin{align*}	
  \overbrace{%
    \{1,\ldots,p\}  \xrightarrow[\text{fit model}]
    {\text{step \rom{1}\ } (\D_1)} 
    \Sall \xrightarrow[\text{select model}]{\text{step \rom{2}\ } (\D_1,\D_2)}
  }^{\text{screening stage}} 
  \Sest 
  \overbrace{\xrightarrow[\text{test support}]{\text{step \rom{3}\ } (\D_3)} \Shat}^{\text{cleaning stage}}
  \enspace.
\end{align*}

Under  suitable conditions,  the screen  and clean  procedure performs
consistent variable selection, that is, it asymptotically recovers the true support with probability one.
The two main assumptions are that the screening stage should asymptotically avoid false negatives, and that the size of the true support should be constant, while the number of candidate variables is allowed to grow logarithmically in the number of examples. 
These assumptions are brought back by \citet{Meinshausen09} in more rigorous
terms as the ``screening property'' and ``sparsity property''.

Empirically, \citet{Wasserman09} tested the procedure with the Lasso, univariate
testing, and forward stepwise regression at step \rom{1} of the screening stage. 
At step \rom{2}, model selection was always based on ordinary least squares (OLS) regression. The OLS parameters were adjusted on the ``training'' subsample $\D_1$, using the variables in $\{\Supp_{\lambda}\}_{\lambda\in\Lambda}$,
and model selection consisted in minimizing the empirical error on the
``validation'' subsample $\D_2$ with respect to $\lambda$.
Cleaning was then finally performed by testing the nullity of the OLS coefficients using the independent ``test'' subsample $\D_3$.
\citet{Wasserman09} conclude that the variants using multivariate regression
(Lasso and forward stepwise) have similar performances, way above univariate
testing.

We now introduce the improvements that we propose here at each stage
of the process.
Our methodological contribution lies at the cleaning stage, but we also
introduced minor modifications at the screening stage that have considerable
practical outcomes.


\subsection{Adaptive-Ridge Cleaning Stage}

The original cleaning stage of \citet{Wasserman09} is based on the ordinary
least square (OLS) estimate. 
This choice is amenable to efficient exact testing procedure for selecting the 
relevant variables, where the false discovery rate can be provably controlled.
However, this advantage comes at a high price:
\begin{itemize}
  \item
  first, the procedure can only be used if the OLS is applicable, which requires that the
  number of variables $\left|\Sest\right|$ that passed the screening stage is
  smaller than the number of examples $\left|\D_3\right|$ reserved for the
  cleaning stage;
  \item
  second, the only information retained from the screening stage is the support 
  $\Sest$ itself.  There are no other statistics about the estimated regression
  coefficients that are transferred to this stage.
\end{itemize}

We propose to make a more effective use of the data reserved for the
screening stage by following the approach described in
Section~\ref{sec:adaptiveridge}: the magnitude of the regression coefficients
$\hatbfbeta{(\hat{\lambda})}$ obtained at the screening stage is 
transferred to the cleaning stage via the
adaptive-ridge penalty term.
Adaptive refers here to the adaptation of the penalty term to the data at hand.
The penalty metric is adjusted to the ``training'' subsample $\D_1$, its strength
is set thanks to the ``validation'' subsample $\D_2$, and 
cleaning is eventually performed by testing the nullity of the adaptive-ridge
coefficients using the independent ``test'' subsample~$\D_{3}$.  

The statistics computed from our penalized cleaning stage improve the power of
the procedure: we observe a dramatic increase in sensitivity (that is, in true
positives) at any  false discovery rate (see Figure~\ref{fig:FDRSENcurves}
of the numerical experiment section). 
With this improved accuracy also comes more precision: the penalization at the
cleaning stage brings the additional benefit of stabilizing the selection 
procedure, with less variability in sensitivity and false discovery rate.
Furthermore, our procedure allows for a cleaning stage remaining in the
high-dimensional setup (that
is, $\left|\Sest\right| \gg \left|\D_3\right|$). 

However, using penalized estimators raises a difficulty for the calibration of 
the statistical tests derived from these statistics. 
We resolved this issue through the use of permutation tests.

\subsection{Testing the Significance of the Adaptive-Ridge Coefficients}
\label{sec:permutationtest}

Student's $t$-test and Fisher's $F$-test are two standard ways of testing the
nullity of the OLS coefficients.
However, these tests do not apply to ridge regression, for which no exact
procedure exists.

\cite{Halawa99} proposed a non-exact $t$-test, but it can be severely off when
the explanatory variables are strongly correlated.
For example, \cite{Cule11} report a false positive rate as high as $32\%$ for a
significance level supposedly fixed at $5\%$.
Typically, the inaccuracy aggravates with high penalty parameters, due to
the bias of the ridge regression estimate, and due to the dependency between the
response variable and the ridge regression residuals.

The $F$-test compares the goodness-of-fit of two nested models.
Let $\hat{\by}_{1}$ and $\hat{\by}_{0}$ be the $n$-dimensional vectors of
predictions for the larger and smaller model respectively.
The $F$-statistic 
\begin{equation}\label{eq:F-statistic}
  F = \frac{\left\|\by - \hat{\by}_0\right\|^2 - \left\|\by - \hat{\by}_1\right\|^2}
            {\left\|\by - \hat{\by}_{1}\right\|^2}
   \enspace,
\end{equation}
follows a Fisher distribution when $\hat{\by}_{1}$ and $\hat{\by}_0$ are
estimated by ordinary least squares under the null hypothesis that the smaller
model is correct.
Although it is widely used for model selection in penalized regression problems
\citep[for calibration and degrees of freedom issues, see][]{Hastie90}, the
$F$-test is not exact for ridge regression, for the reasons already mentioned
above -- estimation bias and dependency between the numerator and the 
denominator in Equation~\eqref{eq:F-statistic}.
Here, we propose to approach the distribution of the $F$-statistic under the
null hypothesis by randomization.
%
We permute the values taken by the explicative variable to be
tested, on the larger model, so as to estimate the distribution of the $F$-statistic under the null
hypothesis that the variable is irrelevant.
This permutation test is asymptotically exact when the tested variable is
independent from the other explicative variables, and is approximate in the
general case.  
It can be efficiently implemented using block-wise decompositions, thereby
saving a factor $p$, as detailed in Appendix~\ref{sec:implementation:details}.
%

\begin{table}
  \caption{Expected false positive rate FPR (or type-\rom{1} error) and 
           sensitivity $\SEN$ (or power) computed over 500 simulations and 
           over the variables selected in the screening stage.
           The prescribed significance level is $5\%$.
           The IND, BLOCK, GROUP and TOEP$^-$ designs are fully described
           in~Section~\ref{sec:simulation}.}
  \label{tab:test:compare}
  \begin{center}
    \begin{tabular}{lrr@{}rrr@{}rrr@{}rrr}
      \hline  \\[-2ex]
      \multirow{2}{*}{Simulation design} & \multicolumn{2}{c}{IND} && \multicolumn{2}{c}{BLOCK} && \multicolumn{2}{c}{GROUP} && \multicolumn{2}{c}{TOEP$^-$} \\[.25ex]
      \cline{2-3} \cline{5-6} \cline{8-9} \cline{11-12} \\[-2ex]
      & \multicolumn{1}{c}{FPR} & \multicolumn{1}{c}{$\SEN$} && \multicolumn{1}{c}{FPR} & \multicolumn{1}{c}{$\SEN$} && \multicolumn{1}{c}{FPR} & \multicolumn{1}{c}{$\SEN$} && \multicolumn{1}{c}{FPR} & \multicolumn{1}{c}{$\SEN$}\\[.25ex]
      \hline \\[-2ex]   
      permutation $F$-test & \color{red}{5.1} & 92.4 && 3.9 & 86.7 && 3.9& 62.3 && 4.7 & 81.9 \\
      standard $F$-test & \color{red}{9.9} & 93.1 && \color{red}{11.8} & 89.6 && \color{red}{14.8} & 73.0 && \color{red}{15.4} & 87.1 \\
      standard $t$-test & \color{red}{8.0} & 94.0 && \color{red}{12.4} & 93.1 && \color{red}{5.8} &95.7 && \color{red}{7.9} & 85.1 \\[.25ex]
      \hline   \\[-4ex]
    \end{tabular}
  \end{center}
\end{table}
Table~\ref{tab:test:compare} shows that, compared to the standard $t$-test and
$F$-test \citep[see e.g.][]{Hastie90}, the permutation test provides a satisfactory control
of the significance level. 
It is either well-calibrated or slightly more conservative than the prescribed
significance level, whereas the standard $t$-test and $F$-test result in false
positive rates that are way above the asserted significance level, especially
for strong correlations between explanatory variables.
These observations apply throughout the experiments reported in
Section~\ref{sec:simulation}.

Testing all variables results in a multiple testing problem. 
We propose here to control the false discovery rate ($\FDR$), which is defined as
the expected proportion of false discoveries among all discoveries.
This control requires to correct the $p$-values for multiple testing
\citep{Benjamini95}. 
The overall procedure is well calibrated as shown in 
Section~\ref{sec:numerical}.  

\subsection{Modifications at Screening Stage}\label{sec:screening}

\cite{Wasserman09} propose to use two subsamples at the cleaning stage in 
order to establish the consistency of the screen and clean procedure.
Indeed, this consistency relies partly on the fact that all relevant variables 
pass the screening stage with very high probability.
This ``screening property'' \citep{Meinshausen09} was established using the
protocol described in Section~\ref{sec:screen_clean:original}.
To our knowledge, it remains to be proved for model selection based on 
cross-validation.
However, \cite{Wasserman09} mention another procedure relying on two
independent subsamples of the original data $\D = \D_1\cup \D_{2}$, where model
selection relies on leave-one-out cross-validation on $\D_1$ and $\D_2$ is 
reserved for cleaning.
The following chart summarizes this modified procedure:
\begin{align*}	
  \overbrace{%
    \{1,\ldots,p\}  \xrightarrow[\text{fit model}]
    {\text{step \rom{1}\ } (\D_1)} 
    \Sall \xrightarrow[\text{select model}]{\text{step \rom{2}\ } (\D_1)}
  }^{\text{screening stage}} 
  \Sest 
  \overbrace{\xrightarrow[\text{test support}]{\text{step \rom{3}\ } (\D_2)} \Shat}^{\text{cleaning stage}}
  \enspace.
\end{align*}
Hence, half of the data are now devoted to each stage of the method.
We followed here this variant, which results in important sensitivity gains for
the overall selection procedure.

We slightly depart from \citep{Wasserman09}, by selecting the model by 10-fold cross-validation, and, more importantly, by using the 
sum of squares residuals of the \emph{penalized} estimator for model 
selection.
Note that \cite{Wasserman09}, and later \cite{Meinshausen09} based model selection on the
OLS estimate using the support $\Supp_{\lambda}$. 
This choice implicitly limits the size of the selected
support $|\Sest| < \tfrac{n}{2}$, which is actually implemented more 
stringently as $|\Sest| \leq \sqrt{n}$ and $|\Sest| \leq \tfrac{n}{6}$ by
\cite{Wasserman09} and \cite{Meinshausen09} respectively.
Our model selection criterion allows for more variables to be transferred to the 
cleaning stage, so that the screening property is more likely to hold.

%
%

\section{Numerical Experiments}
\label{sec:numerical}

Variable selection algorithms are difficult to assess objectively on real data,
where the truly relevant variables are unknown.
Simulated data provide a direct access to the ground truth, in a situation where
the statistical hypotheses hold.
\label{sec:simulation}

\subsection{Simulation Models}

We consider the linear regression model 
%
$
Y = X \bfbeta^\star + \varepsilon
$,
%
where $Y$ is a continuous response variable, $X=(X_1,\dots,X_p)$ is a vector of
$p$ predictor variables, $\bfbeta^\star$ is the vector of unknown parameters and
$\varepsilon$ is a zero-mean Gaussian error variable with variance $\sigma^2$.
The parameter $\bfbeta^\star$ is sparse, that is, the support set 
$\Sreal=\left\{j\in\{1,...,p\}|\betajreal\neq0\right\}$ indexing its 
non-zero coefficients is small $|\Sreal| \ll p$.

Variable selection is known to be affected by numerous factors: the number of
examples $n$, the number of variables $p$, the sparseness of the model
$|\Sreal|$, the correlation structure of the explicative variables, 
the relative magnitude of the relevant parameters 
$\{\betajreal\}_{j\in\Sreal}$, and the signal-to-noise ratio $\SNR$.

In our experiments, we varied $n\in\{250,500\}$, $p\in\{250,500\}$,
$|\Sreal|\in\{25, 50\}$, $\rho \in\{0.5,0.8\}$.
We  considered four predictor correlation structures:
\begin{list}{}{\settowidth\labelwidth{MMMMX}\leftmargin\labelwidth}
  \item[IND] independent explicative variables following a zero-mean, 
  unit-variance Gaussian distribution: $X \sim \mathcal{N} (\bf{0}, \bf{I})$; 
  \item[BLOCK] dependent explicative variables following a zero-mean Gaussian
  distribution, with a block-diagonal covariance matrix: 
  $X \sim \mathcal{N} (\bf{0}, \boldsymbol{\Sigma})$,
  where $\Sigma_{ii} = 1$, $\Sigma_{ij} = \rho$ for all pairs $(i,j),\ j\neq i$
  belonging to the same block and $\Sigma_{ij} = 0$ for all pairs $(i,j)$
  belonging to different blocks.  Each block comprises 25 variables. 

  The position of relevant variables is dissociated from the block structure,
  that is, randomly distributed in $\{1, ..., p\}$.  This design is thus 
  difficult for variable selection.
  \item[GROUP] same as BLOCK, except that the relevant variables are gathered a
  single block when $|\Sreal|=25$ and in two blocks when $|\Sreal|=50$, thus 
  facilitating group variable selection.
  \item[TOEP$^-$] same as GROUP, except that 
  $\Sigma_{ij} = -\rho^{|i - j|}$ for all pairs $(i,j),\ j\neq i$
  belonging to the same block and $\Sigma_{ij} = 0$ for all pairs $(i,j)$
  belonging to different blocks.

  This design allows for strong negative correlations.
\end{list}
The non-zero parameters $\betajreal$ are drawn from a uniform
distribution $\mathcal{U}(10^{-1},1)$ to enable different magnitudes.
Finally, the signal to noise ratio,
defined as $\SNR = {{\bfbeta^\star}^\top \bfSigma\bfbeta^\star}/{\sigma^2}$
varies in $\{4,8,32\}$.

\subsection{Two-Stage Estimation}

In the following, we discuss the IND BLOCK, GROUP and TOEP$^-$ designs with
$n=250$, $p=500$, $|\Sreal|=50$ and $\rho=0.5$.
We report results with three different noise levels.
The relative behavior of the estimation methods is similar for high and medium 
noise levels (respectively $\SNR=4$ and $\SNR=8$), with more significant
differences for medium noise levels. 
The situation then drastically changes for the low noise level ($\SNR=32$).

\begin{figure}[t!]
  \begin{center}
    \begin{tabular}{@{}c@{~}c@{\hspace*{0.05\columnwidth}}c@{~}c@{\hspace*{0.05\columnwidth}}c@{~}c@{}c@{~}c@{\hspace*{0.05\columnwidth}}}
      & \hspace*{0.02\columnwidth}{SNR = 4}& & \hspace*{0.02\columnwidth}{SNR = 8}& & \hspace*{0.02\columnwidth}{SNR = 32} \\
      \rotatebox{90}{\hspace*{0.1\columnwidth}IND} &
      \includegraphics[width=0.275\columnwidth, trim=0.8cm 2cm 1cm 1.5cm, clip=true]{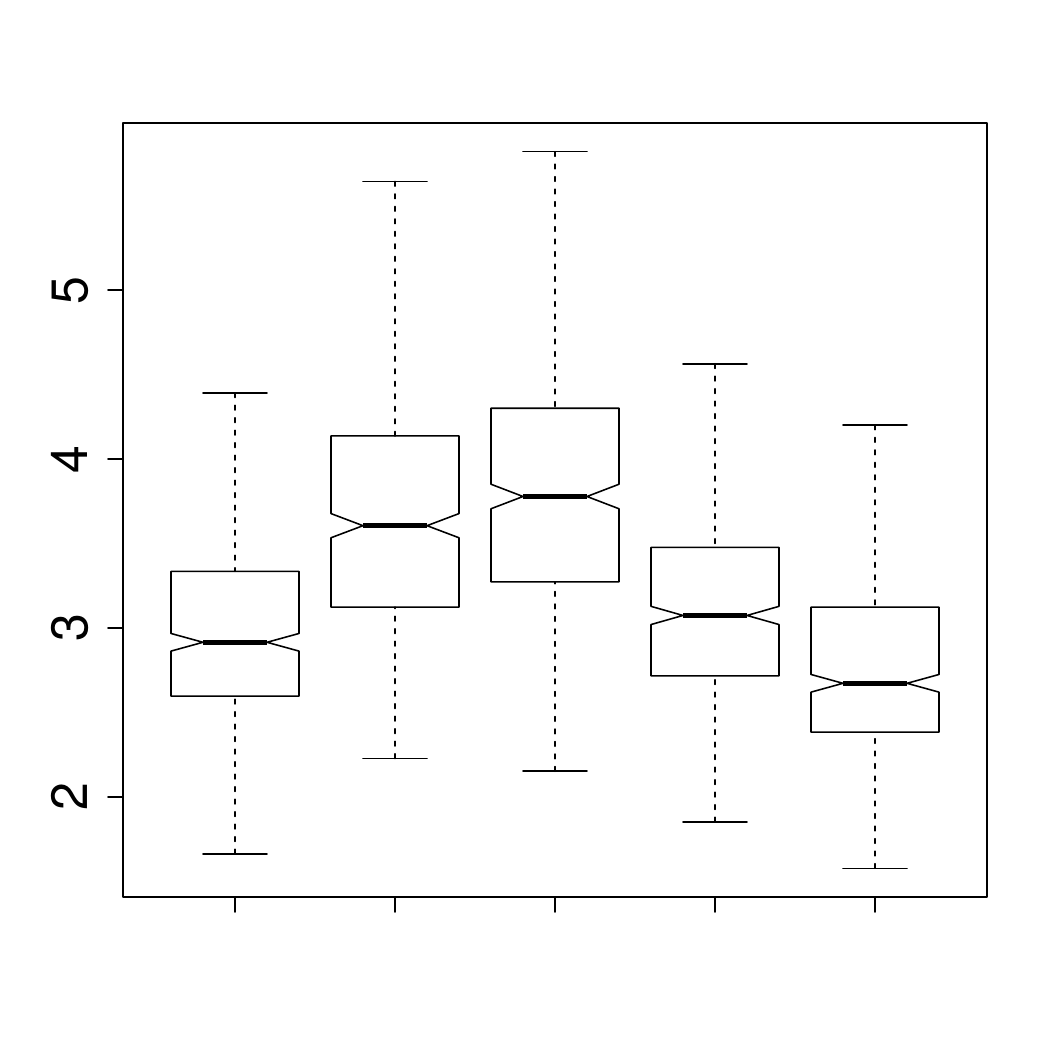} &
       &
      \includegraphics[width=0.275\columnwidth, trim=0.8cm 2cm 1cm 1.5cm, clip=true]{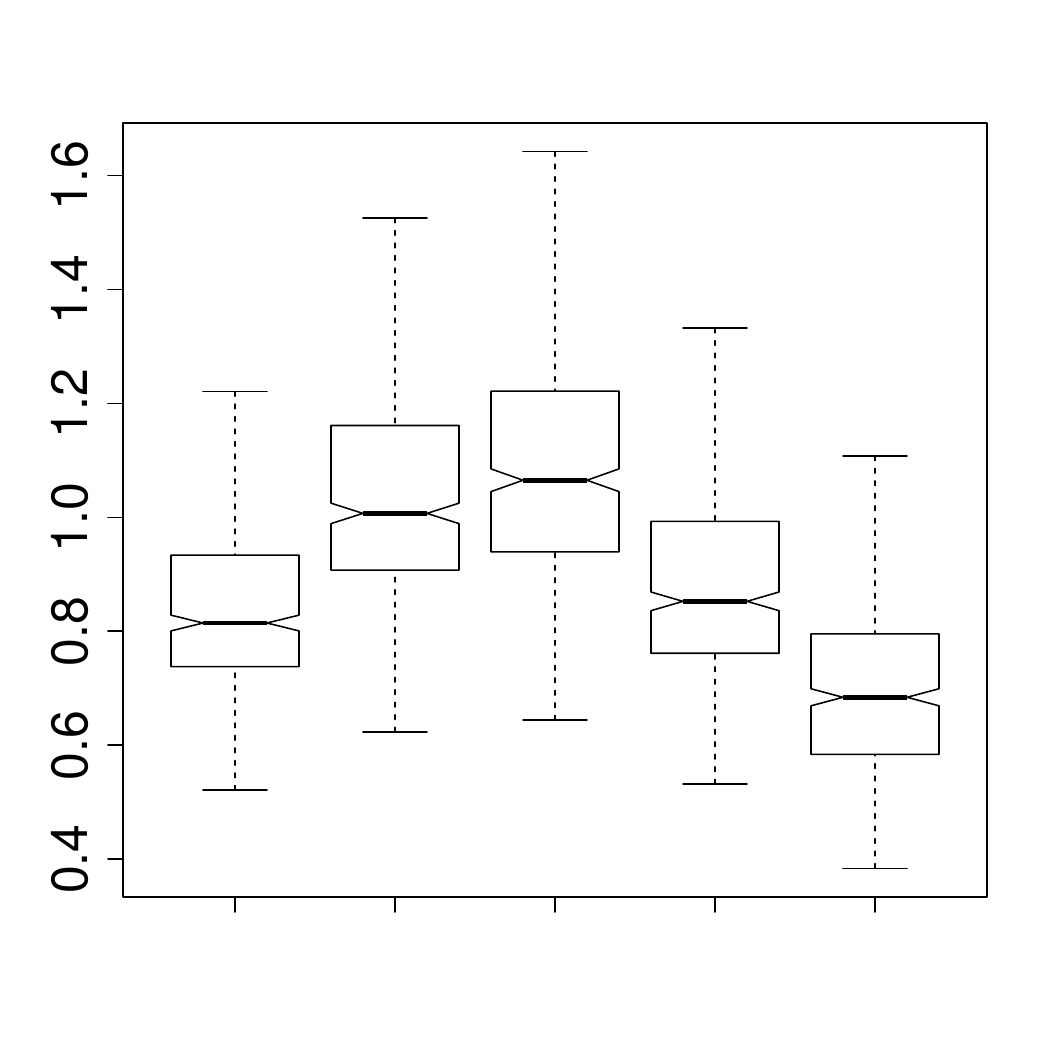}&
       &
      \includegraphics[width=0.275\columnwidth, trim=0.8cm 2cm 1cm 1.5cm, clip=true]{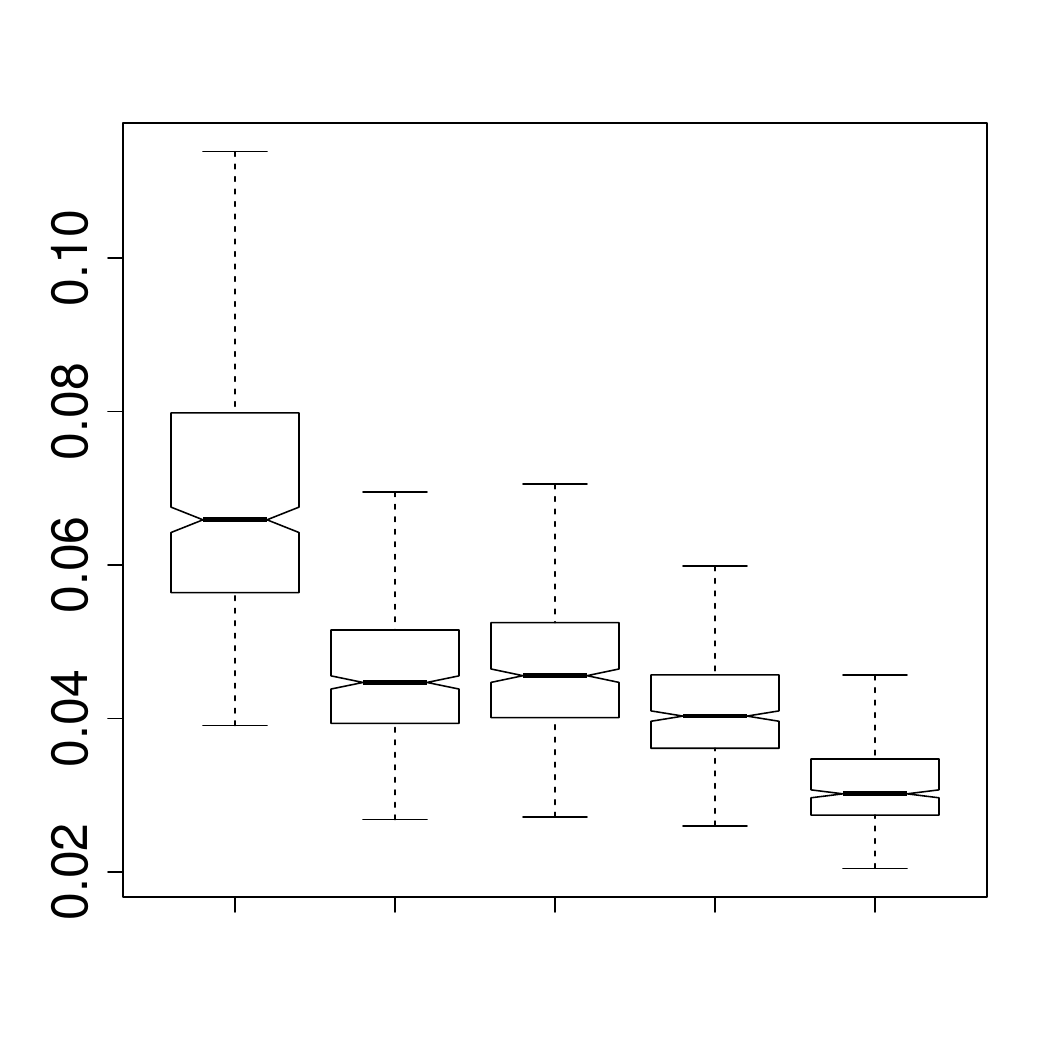}  \\[1ex]
      \rotatebox{90}{\hspace*{0.05\columnwidth}BLOCK} &
      \includegraphics[width=0.275\columnwidth, trim=0.8cm 2cm 1cm 1.5cm, clip=true]{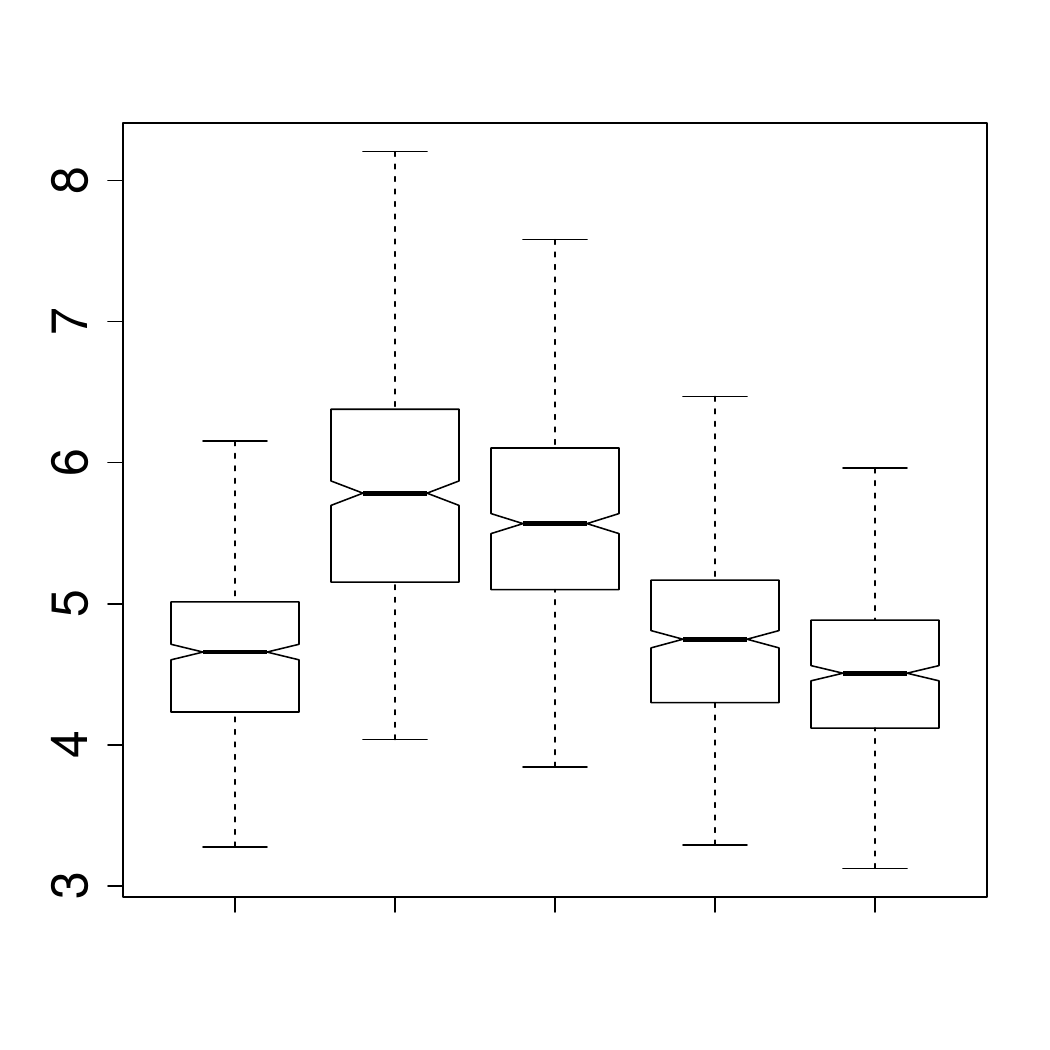} &
       &
      \includegraphics[width=0.275\columnwidth, trim=0.8cm 2cm 1cm 1.5cm, clip=true]{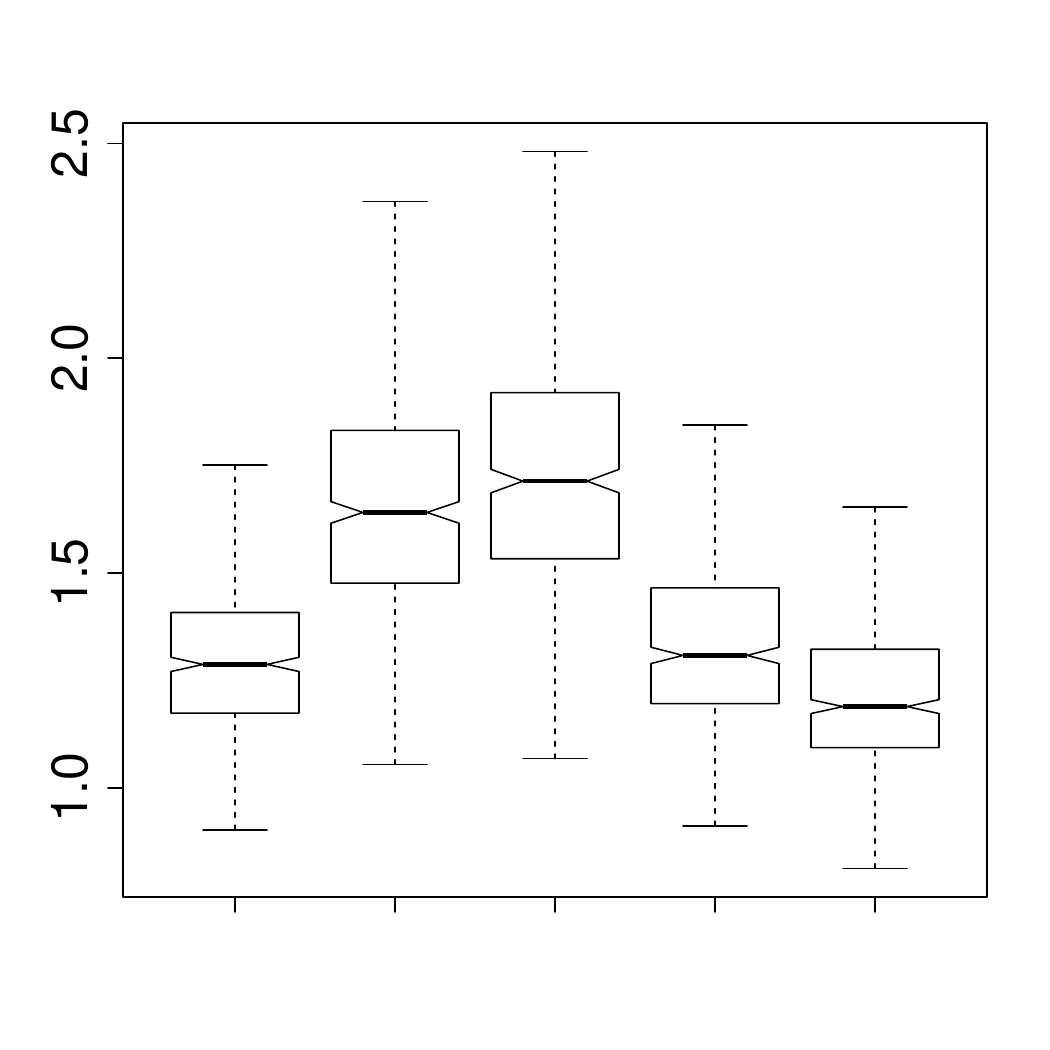}&
       &
      \includegraphics[width=0.275\columnwidth, trim=0.8cm 2cm 1cm 1.5cm, clip=true]{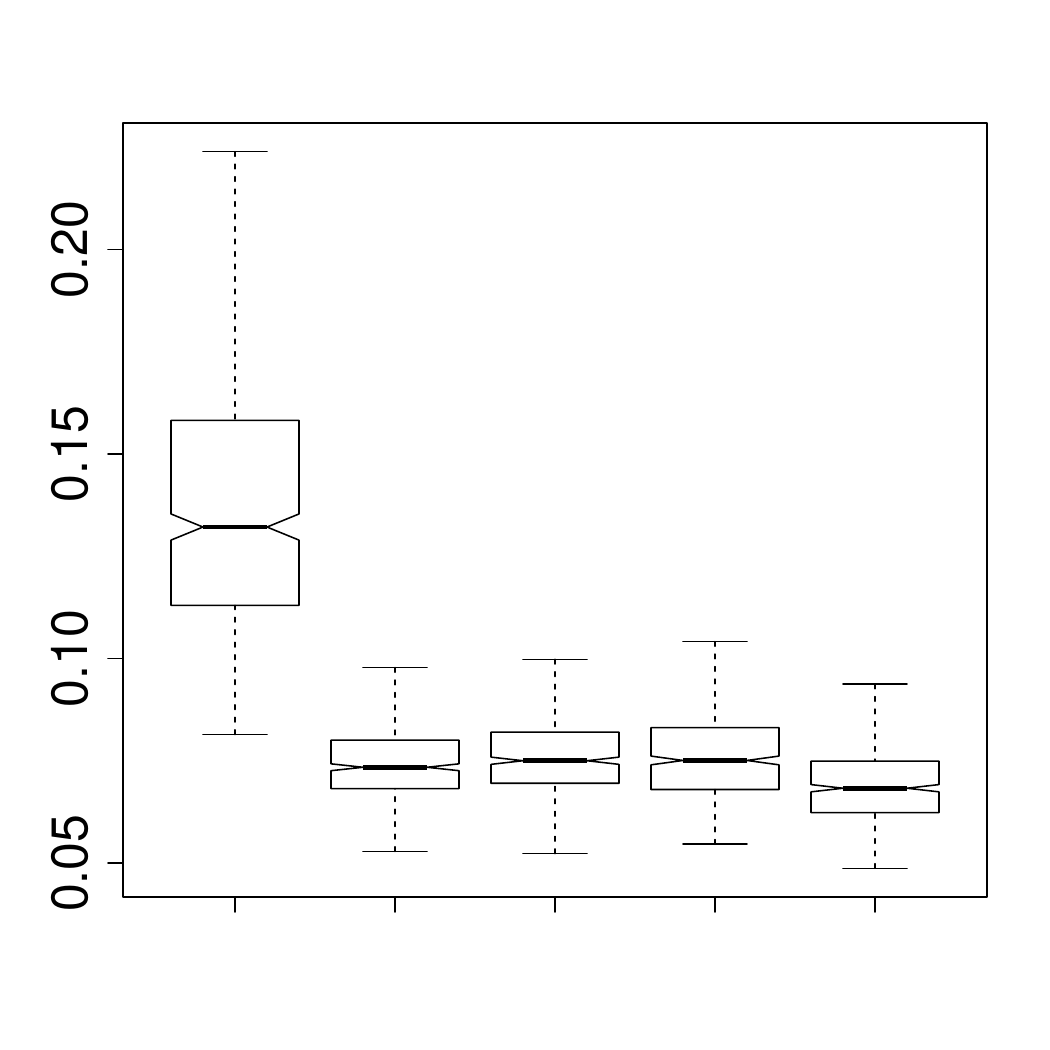}  \\[1ex]
            \rotatebox{90}{\hspace*{0.075\columnwidth}GROUP} &
      \includegraphics[width=0.275\columnwidth, trim=0.8cm 2cm 1cm 1.5cm, clip=true]{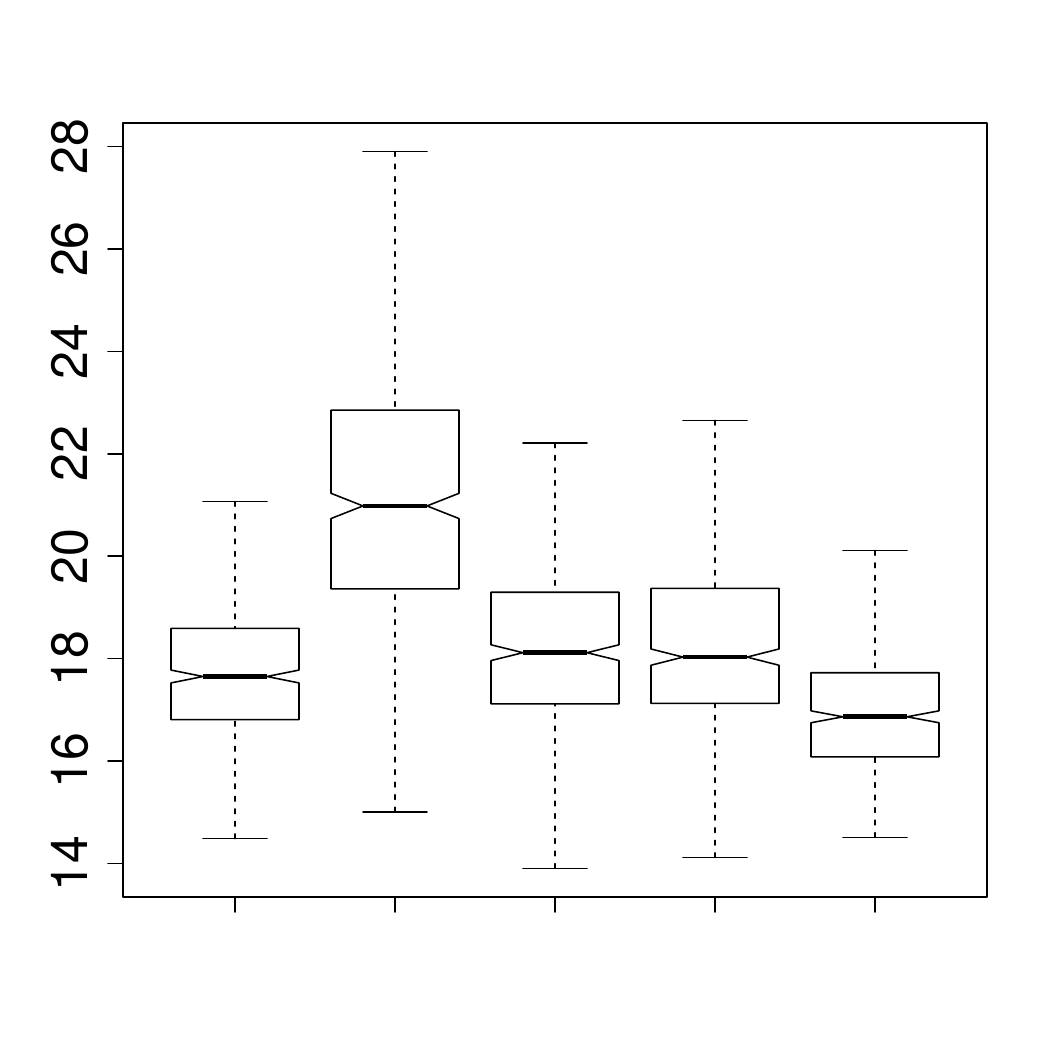} &
       &
      \includegraphics[width=0.275\columnwidth, trim=0.8cm 2cm 1cm 1.5cm, clip=true]{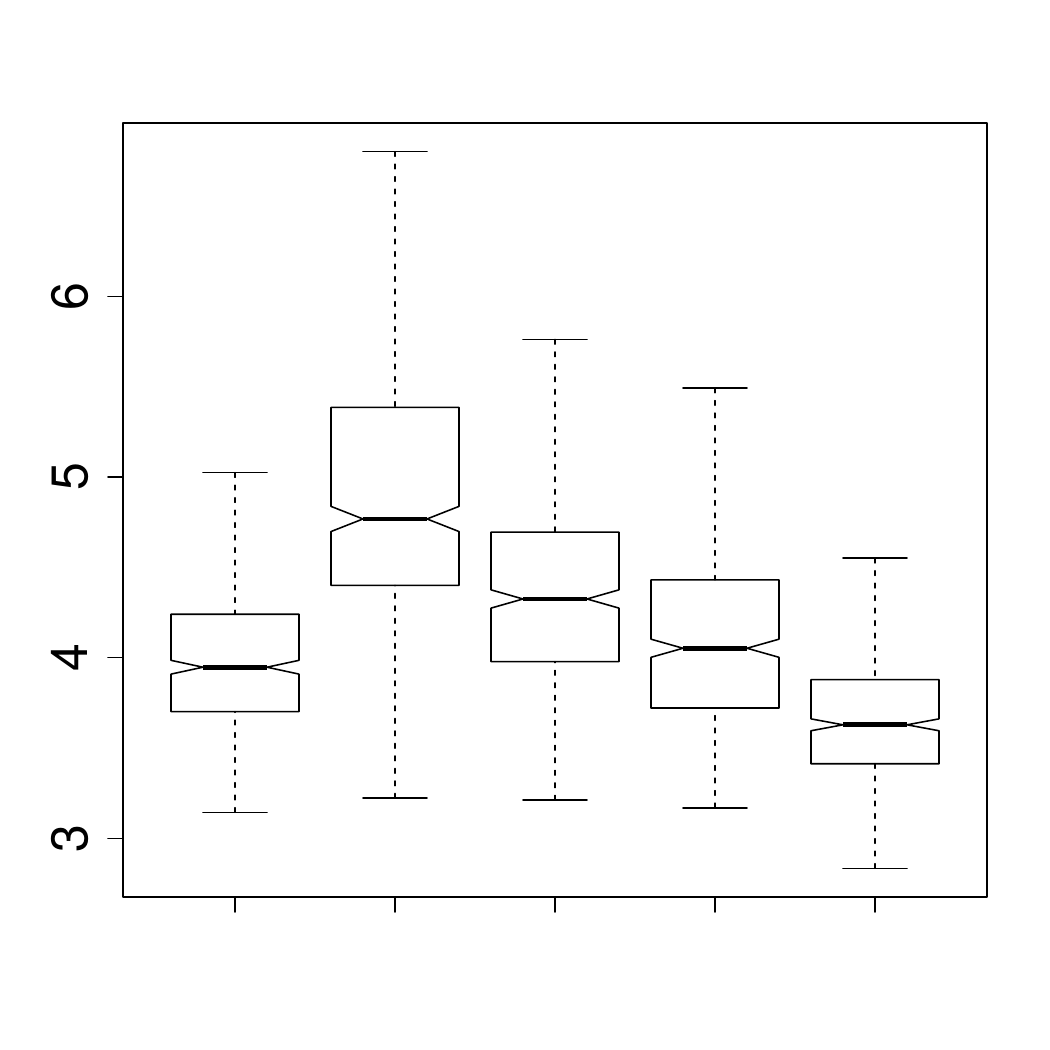}&
       &
      \includegraphics[width=0.275\columnwidth, trim=0.8cm 2cm 1cm 1.5cm, clip=true]{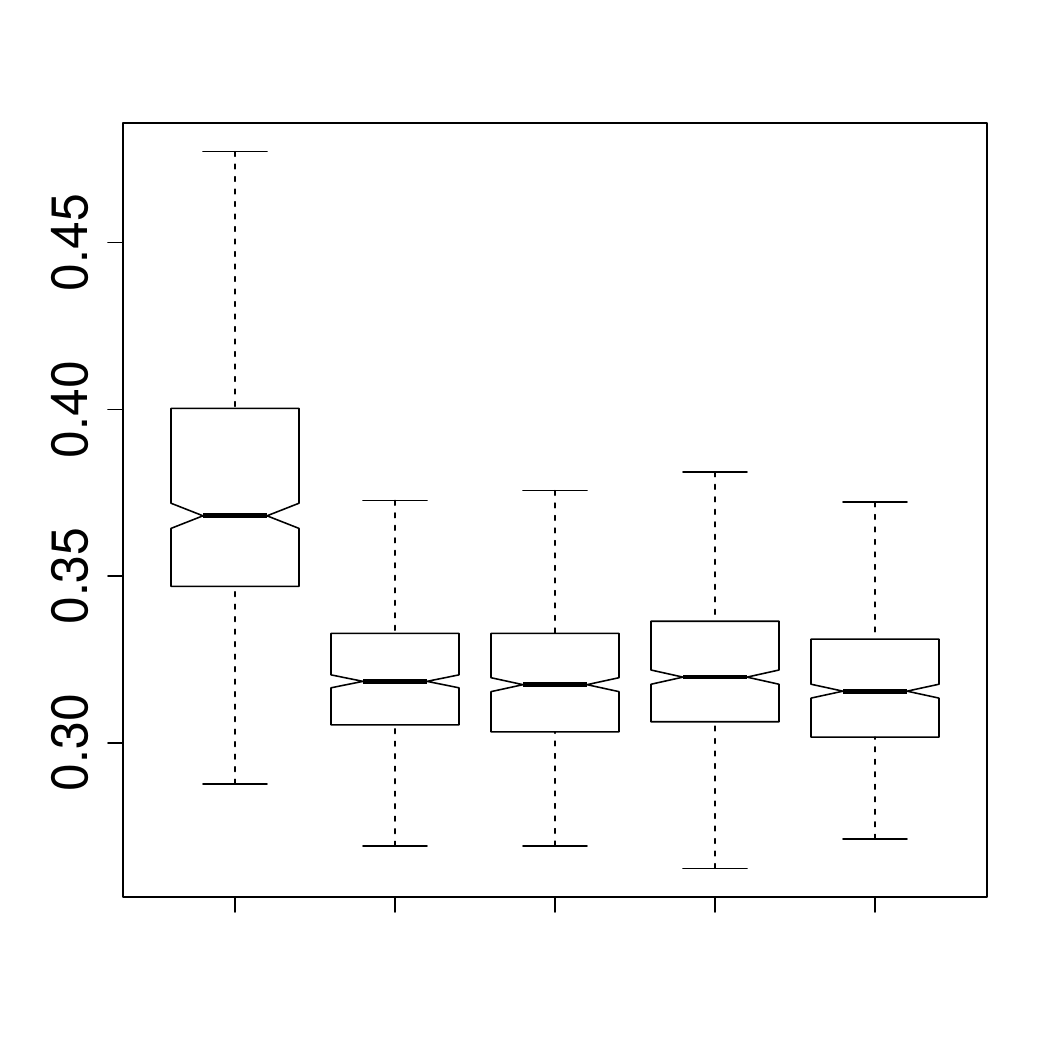}  \\[1ex]
             \rotatebox{90}{\hspace*{0.075\columnwidth}TOEP$^-$} &
      \includegraphics[width=0.275\columnwidth, trim=0.8cm 2cm 1cm 1.5cm, clip=true]{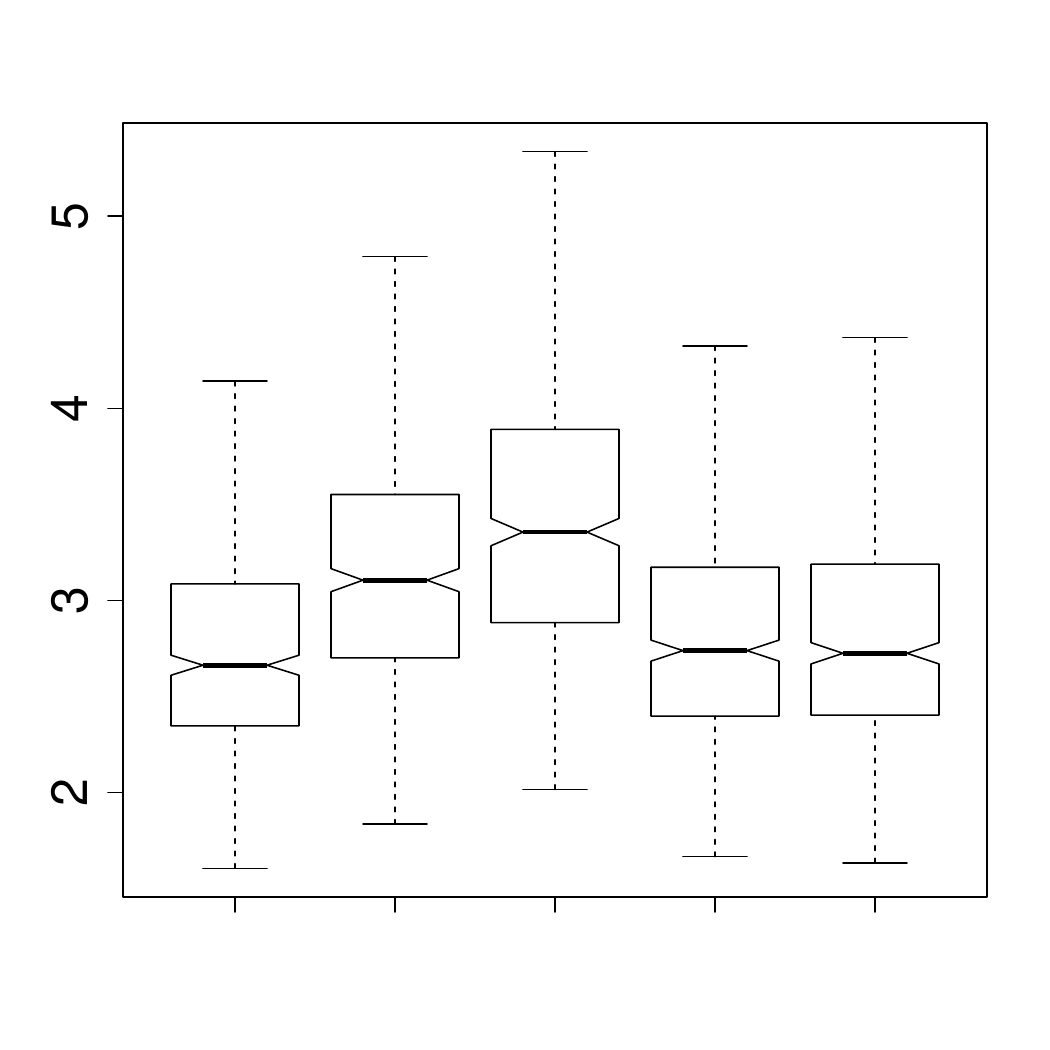} &
       &
      \includegraphics[width=0.275\columnwidth, trim=0.8cm 2cm 1cm 1.5cm, clip=true]{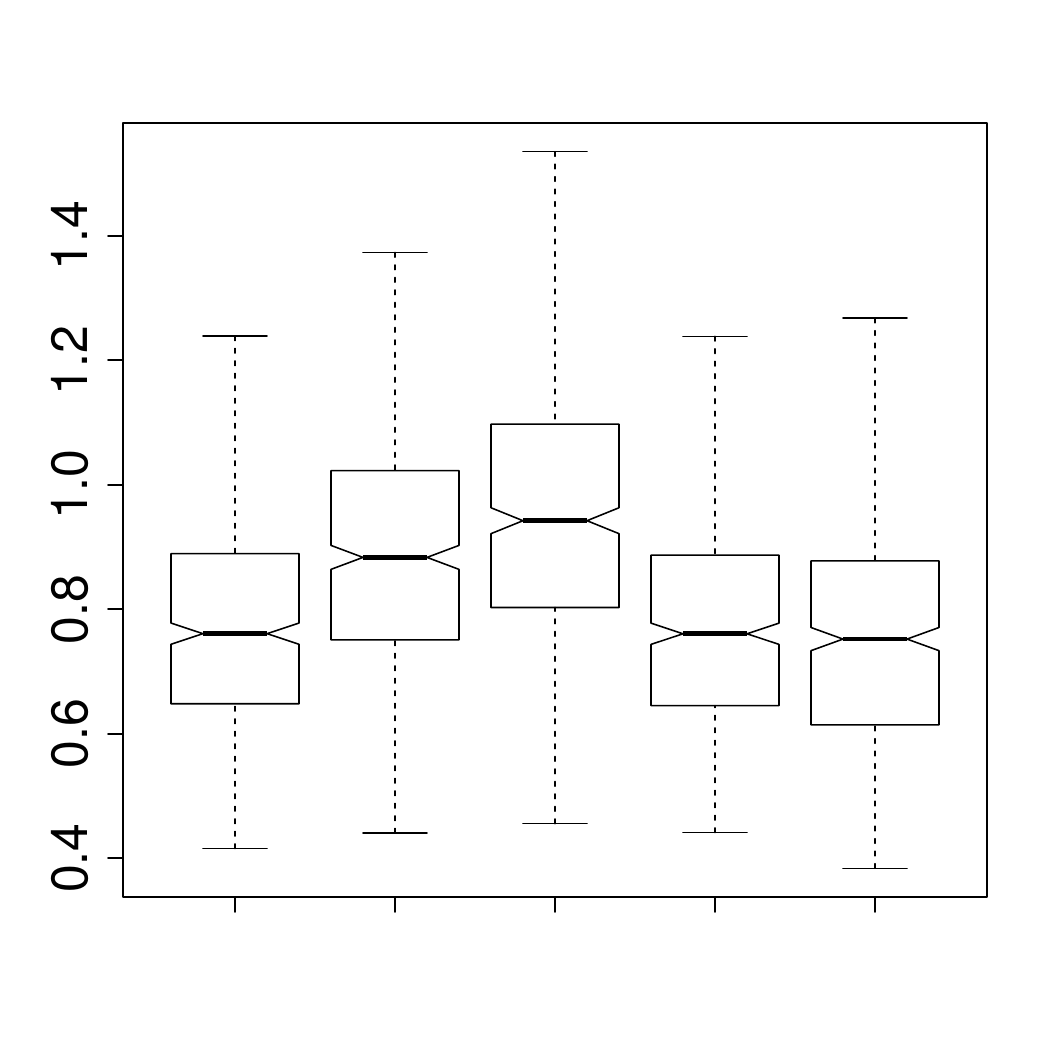}&
       &
      \includegraphics[width=0.275\columnwidth, trim=0.8cm 2cm 1cm 1.5cm, clip=true]{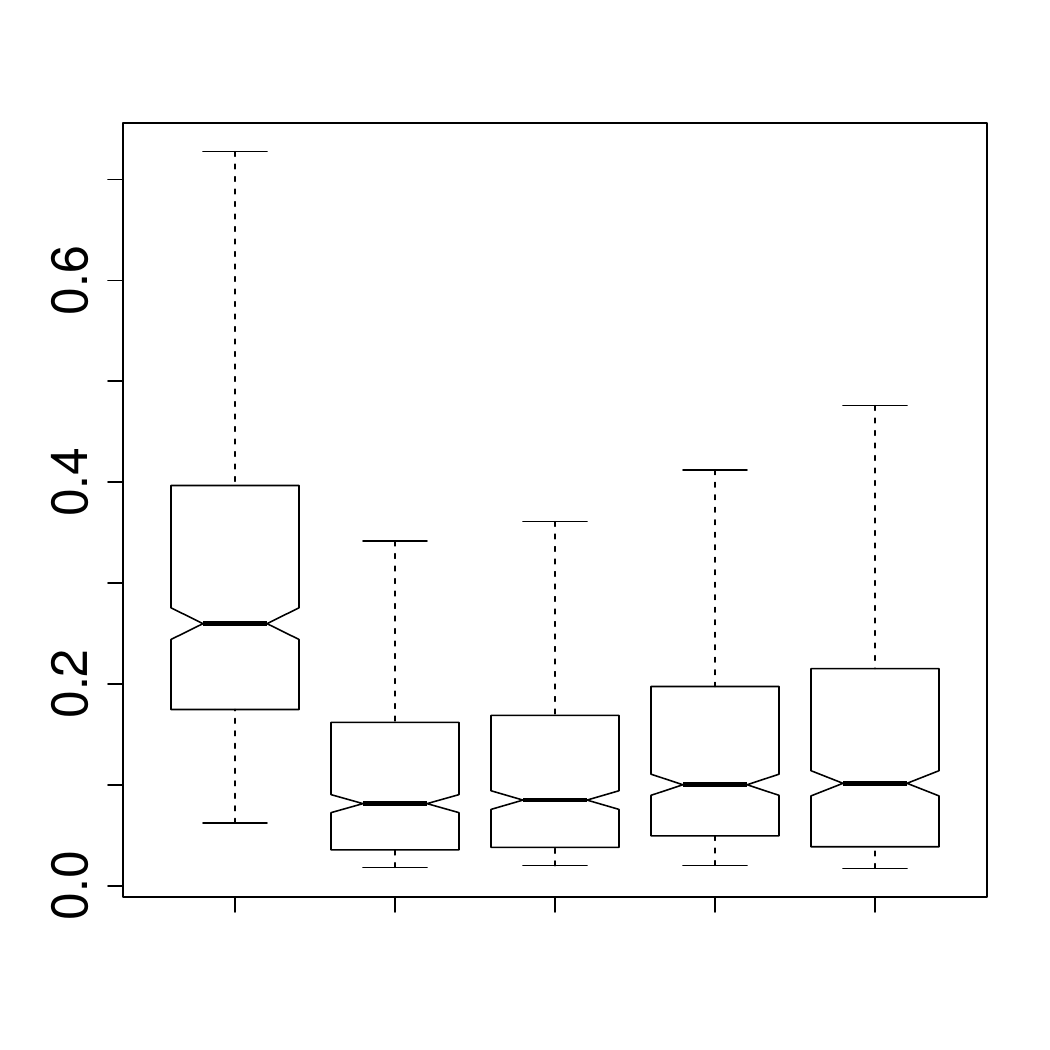}  \\[-1ex]
      &
      \hspace*{0.05\columnwidth}\makebox[0.235\columnwidth][s]{%
      \rotatebox[origin=c]{90}{\scriptsize{L}} 
      \rotatebox[origin=c]{90}{\scriptsize{L+O}} 
      \rotatebox[origin=c]{90}{\scriptsize{L+R}} 
      \rotatebox[origin=c]{90}{\scriptsize{L+A}} 
      \rotatebox[origin=c]{90}{\scriptsize{L\&A}}
      }
      &&
      \hspace*{0.05\columnwidth}\makebox[0.235\columnwidth][s]{%
      \rotatebox[origin=c]{90}{\scriptsize{L}} 
      \rotatebox[origin=c]{90}{\scriptsize{L+O}} 
      \rotatebox[origin=c]{90}{\scriptsize{L+R}} 
      \rotatebox[origin=c]{90}{\scriptsize{L+A}} 
      \rotatebox[origin=c]{90}{\scriptsize{L\&A}}
      }
       &&
      \hspace*{0.05\columnwidth}\makebox[0.235\columnwidth][s]{%
      \rotatebox[origin=c]{90}{\scriptsize{L}} 
      \rotatebox[origin=c]{90}{\scriptsize{L+O}} 
      \rotatebox[origin=c]{90}{\scriptsize{L+R}} 
      \rotatebox[origin=c]{90}{\scriptsize{L+A}} 
      \rotatebox[origin=c]{90}{\scriptsize{L\&A}}
      }
    \end{tabular}
  \end{center}
 \caption{%
  Mean prediction error computed over 500 simulations for each design.
  Lasso direct estimation (L) is compared to: 
  Lasso screening followed by OLS estimation (L+O), 
  Lasso screening followed by ridge estimation (L+R), 
  Lasso screening followed by adaptive-ridge estimation (L+A), 
  jointly optimized Lasso screening with adaptive-ridge estimation (L\&A).
}
  \label{fig:estimation:comparison}
\end{figure}

We compare the variants of the two-stage estimation methods based on the 
predictive mean squared error. Similar conclusions would be drawn from the 
accuracy measures on the vector of coefficients $\bfbeta^\star$.
Figure~\ref{fig:estimation:comparison} displays the boxplots of prediction 
error obtained over 500 simulations for each design.

There is no benefit in a post-Lasso estimation step for high an medium 
noise levels ($\SNR\in \{4,8\}$).
OLS and ridge post-processing then have important detrimental effects 
and adaptive ridge has still a slight unfavorable effect. 
It is only when the two-step procedure is jointly optimized with respect to the 
two penalization parameters (by cross-validation), that some improvements become
visible for the first three setups.

When the signal-to-noise ratio is high ($\SNR=32$), Lasso highly benefits from
the second stage whatever it may be (OLS, ridge or adaptive-ridge).  There is a
slight edge to adaptive ridge when variables are independent, but otherwise all
methods are at par.
Globally, the best option here consists again in jointly optimizing the two
stages with respect to the two penalization parameters; some additional
improvements come into view.

Compared to previous studies, which mainly focused on large sample and/or
low-noise settings, our experiments demonstrate that post-Lasso estimation can
have consequential beneficial or detrimental effects in small sample regression. 
In addition to the experimental design, the results vary also considerably
according to the strategy governing the choice of the penalty parameters.
Other experiments (not shown here) attest that using more stringent screening
stages \citep[using the so-called ``1-SE rule'' of][that chooses the highest 
penalty within one standard deviation of the minimum of cross-validation]{Breiman84} lead to better
post-Lasso estimation in some experimental setups, but this is not systematic: in the 
TOEP$^-$ design, this is by far the least favorable option.
Overall, the joint optimization with respect to the two penalization parameters
seems to be a very challenging contender. 
This is also true when the Lasso screening is followed by OLS or ridge regression.
The joint optimization of penalization parameter favors a stringent Lasso
screening compared to the strategy based on serial cross-validation, and a less
stringent one compared to the 1-SE rule.
Though this solution is the most expensive from the computational viewpoint, it 
seems to be also the most effective one regarding predictive mean squared 
error.

\subsection{Two-Stage Inference}

In the following, we discuss the IND BLOCK, GROUP and TOEP$^-$ designs with $n=250$,
$p=500$, $|\Sreal|=25$, $\rho=0.5$ and $\SNR=4$, since the relative behavior of all
methods is representative of the general pattern that we observed across all
simulation settings.  
These setups lead to feasible but challenging problems for model selection.

All variants of the screen and clean procedure are evaluated here with respect
to their sensitivity ($\SEN$), 
for a controlled false discovery rate $\FDR$.  These two measures are the
analogs of power and significance in the single hypothesis testing framework:
\begin{align*}
  \SEN = \mathbb{E} \left[ \dfrac{TP}{TP + FN} \mathbb{I}_{\{(TP+FN)>0\}} \right] 
  \enspace, \enspace
  \FDR = \mathbb{E}\left[ \dfrac{FP}{TP + FP}  \mathbb{I}_{\{(TP+FP)>0\}} \right] 
  \enspace,
\end{align*}
where $FP$ is the number of false positives, $TP$ is the number of true
positives and $FN$ is the number of false negatives.

We first show the importance of the cleaning stage for $\FDR$ control. 
We then show the benefits of our proposal compared to the original
procedure of \citet{Wasserman09} and to the univariate approach.
The variable selection method of \citet{Lockhart14} was not included in these
experiments, because it did not produce convincing results in these small $n$ 
large $p$ designs where the noise variance
is not assumed to be known.

\paragraph{Importance of the Cleaning Stage}

Table \ref{tab:compare} shows that the cleaning step is essential to control
the $\FDR$ at the desired level.
\begin{table}
  \caption{False discovery rate $\FDR$ and  sensitivity $\SEN$, computed over 500
           simulations for each design.
           The screening stage (before cleaning) is not calibrated;
           the cleaning stage is calibrated to control 
           the $\FDR$ below $5\%$, using the Benjamini-Hochberg procedure.
           Our adaptive-ridge (AR) cleaning is compared with the original (OLS) 
           cleaning
           and univariate testing (Univar).}
  \label{tab:compare}
  \begin{center}
    \begin{tabular}{@{}ll@{}rr@{}r@{}rr@{}r@{}rr@{}r@{}rr@{}}
      \hline  \\[-2ex]
      Simulation design && \multicolumn{2}{c}{IND} && \multicolumn{2}{c}{BLOCK} && \multicolumn{2}{c}{GROUP}  && \multicolumn{2}{c}{TOEP$^-$} \\[.25ex]
      \cline{3-4} \cline{6-7} \cline{9-10}  \cline{12-13} \\[-2ex]
      && $\FDR$ & $\SEN$ &\hspace*{1em}& $\FDR$ & $\SEN$  &\hspace*{1em}& $\FDR$ & $\SEN$ &\hspace*{1em}& $\FDR$ & $\SEN$ \\[.5ex]
      \hline \\[-2ex]   
       Before cleaning && \color{red}{76.7} & 87.5 && \color{red}{76.0} & 83.9 && \color{red}{38.9} & 86.2  && \color{red}{79.9} & 56.5 \\
        AR cleaning     &&  4.2 & 76.1 &&  2.8 & 64.8 &&  1.7 & 37.7 && 4.3 & 39.6 \\
        OLS cleaning    &&  3.9 & 48.3 && 3.1 & 37.1 && 2.5 & 17.9 && 3.7 & 25.3 \\
      {Univar}          && 4.4 & 40.4 && \color{red}{86.4} & 71.0 && \color{red}{5.3} & 100.0 && 4.2&28.4 \\
      \hline   \\[-4ex]
    \end{tabular}
  \end{center}
\end{table}
In the screening stage, the variables selected by the Lasso are way too numerous:
first, the penalty parameter is determined to optimize the cross-validated mean
squared error, which is not optimal for model selection;
second, we are far from the asymptotic regime where support recovery can be
achieved.
As a result, the Lasso performs rather poorly.
Cleaning enables the control of the $\FDR$, leading of course to a decrease in
sensitivity, which is moderate for independent variables, and higher in the
presence of correlations.
%

\paragraph{Comparisons of Controlled Selection Procedures}

\begin{figure}
  \begin{center}%
    \begin{tabular}{@{}c@{\hspace*{1ex}}c@{\hspace*{0.1\columnwidth}}c@{~}c@{}}%
      & \hspace*{0.02\columnwidth}{IND} & & \hspace*{0.02\columnwidth}{BLOCK} \\
      \rotatebox{90}{\hspace*{0.145\columnwidth}$\SEN$} &
      \includegraphics[width=0.35\columnwidth, trim=1.05cm 1.5cm 1cm 2cm, clip=true]{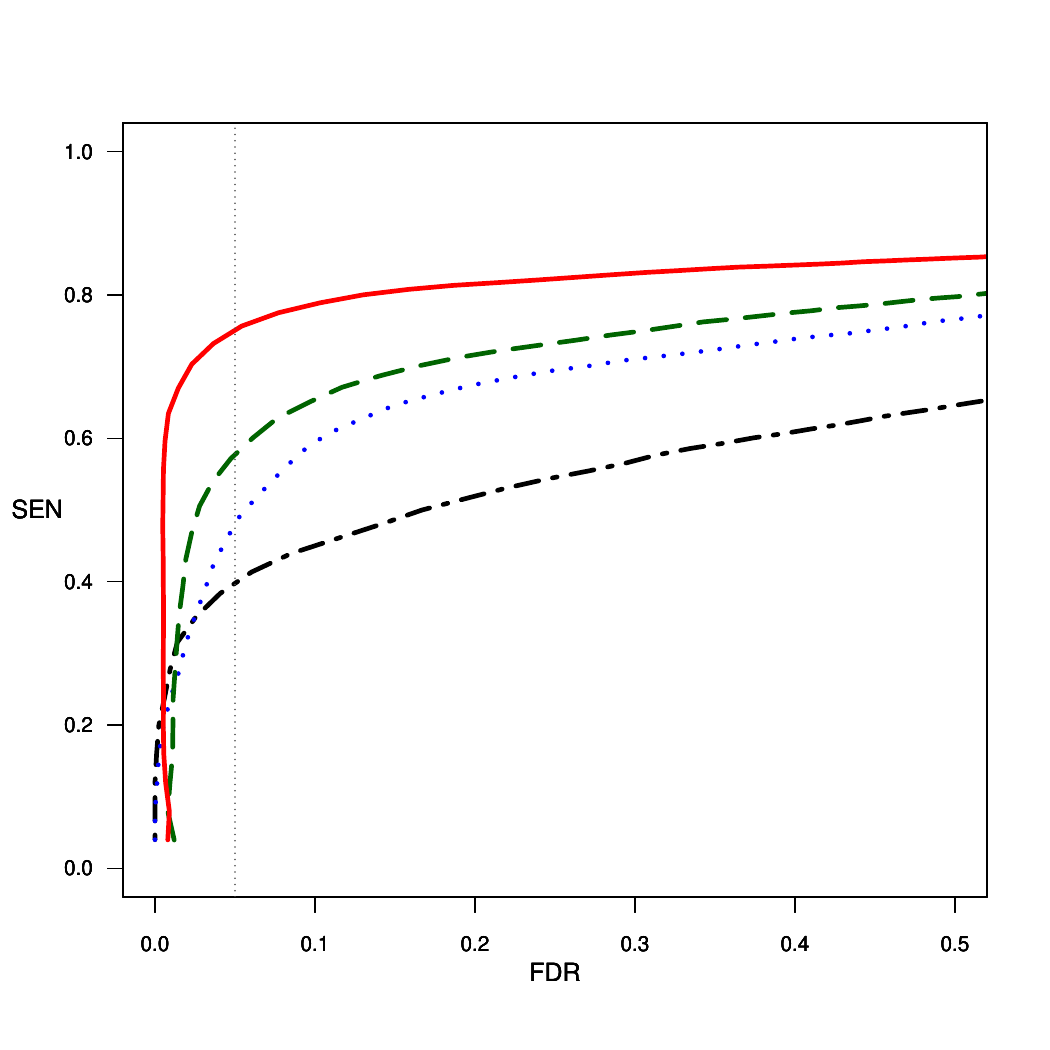} &
      \rotatebox{90}{\hspace*{0.145\columnwidth}$\SEN$} &
      \includegraphics[width=0.35\columnwidth, trim=1.05cm 1.5cm 1cm 2cm, clip=true]{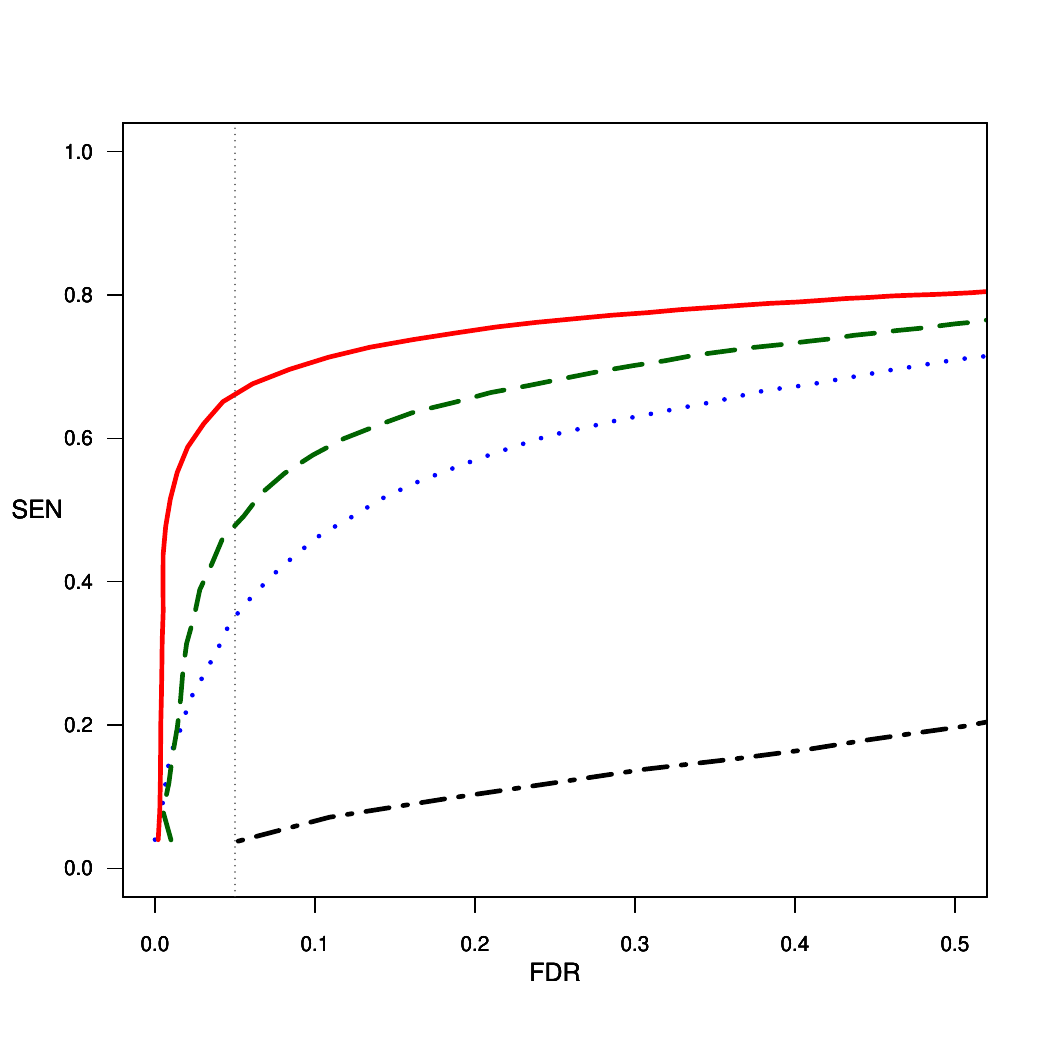} \\[-.75ex]
      & {\hspace*{0.02\columnwidth}\small $\FDR$} & & {\hspace*{0.02\columnwidth}\small $\FDR$} \\[2ex]
      & \hspace*{0.02\columnwidth}{GROUP} & & \hspace*{0.02\columnwidth}{TOEP$^-$} \\
      \rotatebox{90}{\hspace*{0.145\columnwidth}$\SEN$} &
      \includegraphics[width=0.35\columnwidth, trim=1.05cm 1.5cm 1cm 2cm, clip=true]{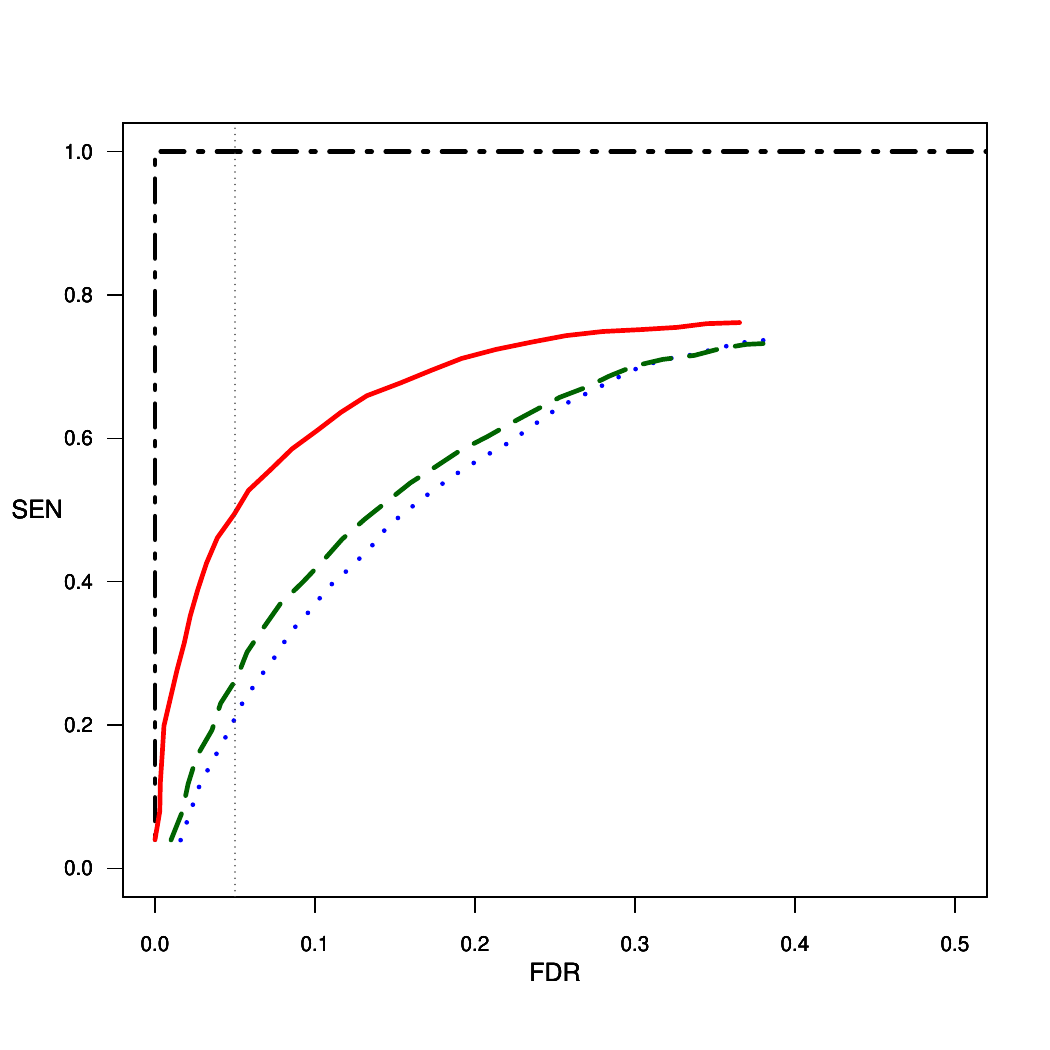} &
      \rotatebox{90}{\hspace*{0.145\columnwidth}$\SEN$} &
      \includegraphics[width=0.35\columnwidth, trim=1.05cm 1.5cm 1cm 2cm, clip=true]{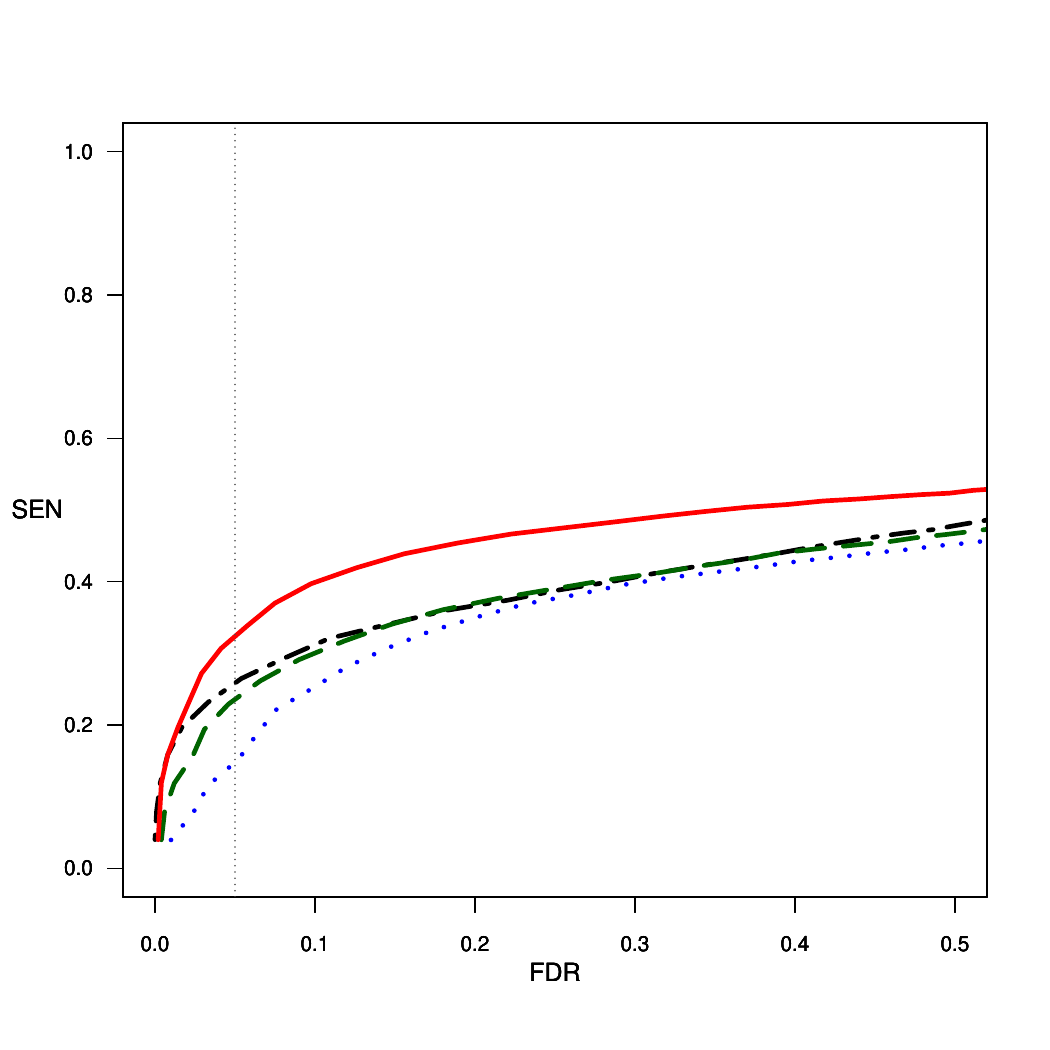} \\[-.75ex]
      & {\hspace*{0.02\columnwidth}\small $\FDR$} & & {\hspace*{0.02\columnwidth}\small $\FDR$}
    \end{tabular}
  \end{center}
  \caption{Sensitivity $\SEN$ versus False Discovery Rate $\FDR$ (the higher, the
  better).  Lasso screening followed by: adaptive-ridge cleaning (red solid
  line), ridge cleaning (green dashed line), OLS cleaning (blue dotted line);
  univariate testing (black dot-dashed line).  All curves are indexed by the
  rank of the test statistics, and averaged over the $500$ simulations of each
  design.
  The vertical dotted line marks the 5\% $\FDR$ level.}
  \label{fig:FDRSENcurves}
\end{figure}

Figure \ref{fig:FDRSENcurves} provides a global picture of sensitivity 
according to $\FDR$, for the test statistics computed in the cleaning stage.
First, we observe that the direct univariate approach, which simply considers a
$t$-statistic for each variable independently, is by far the worst option in the 
IND, BLOCK and TOEP$^-$ designs, and by far the best in the GROUP design.
In this last situation, the univariate approach confidently detects all the
correlated variables of the relevant group, while the regression-based
approaches are hindered by the high level of correlation between variables.
Betting on the univariate approach may thus be profitable, but it is also 
risky due to its extremely erratic behavior.
Second, we see that our adaptive-ridge cleaning systematically dominates the
original OLS cleaning.
To isolate the effect of transfering the magnitude of weights from the effect of
the regularization brought by adaptive-ridge, we show the results obtained from
a cleaning step based on plain ridge regression (with regularization parameter
set by cross-validation).
We see that ridge regression cleaning improves upon OLS cleaning, but that 
adaptive-ridge cleaning brings this improvement much further, thus confirming 
the value of the weight transfer from the screening stage to the cleaning stage.


Table \ref{tab:compare} shows the actual operating conditions of the  
variable selection procedures, when a threshold on the test statistics has to 
be set to control the $\FDR$.  
Here, the threshold is set  to control the $\FDR$ at a $5\%$ level, using a
Benjamini-Hochberg correction.  
This control is always effective for the screen and clean procedures, but not 
for variable selection based on univariate testing.  
In all designs, our proposal dramatically improves over the original OLS
strategy, with sensitivity gains ranging from
$50\%$ to $100\%$.
All differences in sensitivity are statistically significant. 
The variability of $\FDR$ and sensitivity is not  shown to avoid clutter, but 
the smallest variability in $\FDR$ is obtained for the adaptive-ridge cleaning, 
while the smallest variability in sensitivity is obtained for univariate regression, 
followed by adaptive-ridge cleaning. The adaptive ridge penalty thus brings 
about more stability to the overall selection process.

\section{Discussion}

We propose to use the magnitude of regression coefficients in two-stage 
variable selection procedures.
First, we use the connection between the Lasso and 
adaptive-ridge \citep{Grandvalet98a} to convey more information from the 
screening stage to the second stage:
the magnitude of the coefficients estimated at the screening stage is 
transferred to the second stage through penalty parameters.
%

Empirically, our procedure brings marginal improvements when the second stage
aims at improving the regression coefficients \citep{Belloni13,Liu14}, and it
provides sensible improvements compared to the original screen and clean
procedure \citep{Wasserman09} when assessing the uncertainties pertaining to the
selection of relevant variables.
In the first setup, screening and estimation are performed on the same data set,
whereas in the second one, the screening and cleaning stages operate on two
distinct subsamples of data: the transfer is more valuable in this situation.

Regarding post-Lasso estimation, our experiments demonstrate that two-stage
methods can have consequential beneficial or detrimental effects in small sample
regression.
The results vary considerably according to the strategy governing the choice of
the penalty parameters, but the joint optimization with respect to the two
penalization parameters is the most effective one regarding predictive mean
squared error.

For screen and clean, we obtained 
a better control of the False Discovery Rate, which extends to more
difficult settings, with high correlations between variables.
Furthermore, the sensitivity obtained by our cleaning stage is always as good,
and often much better than the one based on the ordinary least squares.
The penalized second step also allows for a less severe screening, since
the second stage can then handle more than $n/2$ variables.
Our procedure can thus be employed in very high-dimensional settings, as the
screening property (that is, in the words of \cite{Buhlman13}, the ability of
the Lasso to select all relevant variables) is more easily fulfilled, which is
essential for a reliable control of the false discovery rate.

Several interesting directions are left for future works. 
The second stage can accommodate arbitrary penalties, and our
efficient implementation applies to all penalties for which a quadratic
variational formulation can be derived.  This is particularly appealing for
structured penalties such as the fused-lasso or the group-Lasso, allowing to use
the knowledge of groups at the second stage, through penalization
coefficients.

On the theoretical side, many interesting issues are raised.  In
particular, we would like to back-up the empirical improvements that have been
almost systematically observed by an apposite analysis.  We believe that two
tracks are promising: first by exploiting that the screening stage transfers a
quantified response to the cleaning stage through the penalization coefficients,
and second, that screening needs not to be stringent, due to the ability of our
second stage to handle more variables.

\section*{Software}

Software and simulations are in the form of R package named ``\if0\blind
{ridgeAdap}\fi '' available on
the personal author page \if0\blind
{(\url{https://www.hds.utc.fr/~becujean})}\fi.

\section*{Acknowlegements}
\if0\blind
{
This work was supported by the UTC foundation for Innovation, in the ToxOnChip 
program.
It has been carried out in the framework of the Labex MS2T
(ANR-11-IDEX-0004-02) within the ``Investments for the future'' program, managed
by the National Agency for Research.
}\fi
\bibliography{biblio_pvalue}

\appendix
\section{Variational Equivalence}\label{sec:equivalence}

We show below that the quadratic penalty in $\bfbeta$ in Problem
\eqref{eq:lasso:variational} acts as the Lasso penalty $\lambda
\norm[1]{\bfbeta}$.

\begin{proof}
The Lagrangian of Problem \eqref{eq:lasso:variational}
is:
\begin{equation*}
  L(\bfbeta) = J(\bfbeta) + \lambda \sum_{j=1}^p \frac{1}{\tau_{j}} \beta_{j}^{2}
      + \nu_0 \bigg( \sum_{j=1}^p \tau_j -\norm[1]{\bfbeta} \bigg) 
      - \sum_{j=1}^p \nu_j \tau_j
  \enspace.
\end{equation*}
Thus, the first order optimality conditions for $\tau_j$ are
\begin{align*}
  \frac{\partial L}{\partial \tau_j}(\tau_j^\star)  = 0 
  & \Leftrightarrow
  - \lambda\frac{\beta_{j}^2}{{\tau_j^\star}^2} + \nu_0 -  \nu_j  = 0 \\
  &\Leftrightarrow
  - \lambda\beta_{j}^2 + \nu_0 \,{\tau_j^\star}^2 - \nu_j \,{\tau_j^\star}^2 = 0 \\
  &\Rightarrow
  - \lambda\beta_{j}^2 +  \nu_0 \,{\tau_j^\star}^2 = 0
  \enspace,
\end{align*}
where the term in $\nu_j$ vanishes due to complementary slackness, which implies
here $\nu_j \tau_j^\star = 0$.
Together with the constraints of Problem \eqref{eq:lasso:variational}, the last
equation implies $\tau_j^\star = \left|\beta_{j}\right|$, hence
Problem~\eqref{eq:lasso:variational}
is equivalent to
\begin{equation*}
 \min_{\bfbeta\in\mathbb{R}^p} J (\bfbeta) +
     \lambda \norm[1]{\bfbeta}
  \enspace, \label{eq:PARf} 
\end{equation*}
which is the original Lasso formulation. 
\end{proof}
%

\section{Efficient Implementation}\label{sec:implementation:details}

Permutation tests rely on the simulation of numerous data sampled under the null
hypothesis distribution.
The number of replications must be important to estimate the rather extreme
quantiles we are typically interested in.
Here, we use $B=1000$ replications for the $q=|\Sest|$ variables selected in the
screening stage.
Each replication involving the fitting of a model, the total computational cost
for solving these $B$ systems of size $q$ on the $q$ selected
variables is $O(Bq(q^{3}+q^{2}n))$.
In the situation where $q\ll B$, great computing savings can be obtained 
using block-wise decompositions and inversions.

First, we recall that the adaptive-ridge estimate, computed at the cleaning 
stage, is computed as 
\begin{equation*}
  \hatbfbeta = \left({\bfX}^{\top}\bfX + \bfLambda\right)^{-1} 
                            \bfX^{\top}\bfy
  \enspace,
\end{equation*}
where $\bfLambda$ is the diagonal adaptive-penalty matrix defined at the
screening stage, whose $j$th diagonal entry is ${\lambda}/{\tau_{j}^\star}$, as defined 
in~(\ref{eq:lasso}--\ref{eq:adaptive-ridge_penalty}).

In the $F$-statistic \eqref{eq:F-statistic}, the permutation affects the
calculation of the larger model $\hat{\by}_{1}$, which is denoted
$\hatyhyponepermute$ for the $b$th permutation. 
Using a similar notation convention for the design matrix and the estimated 
parameters, we have  $\hatyhyponepermute=\Xperm \betaperm$.
When testing the relevance of variable $j$, $\Xperm$ is defined as the
concatenation of the permuted variable $\varj$ and the other original variables:
$\Xperm = (\varj, \bx_1, ..., \bx_{j-1}, \bx_{j+1}, ... \bx_{p})$. 
Then, $\betaperm$ can be efficiently computed by using $\aperm\in\mathbb{R}$, $\varnotj\in\mathbb{R}^{q-1}$ and 
$\betanotj\in\mathbb{R}^{q-1}$ defined as follows:
\begin{align*}
  \aperm &=  (\| \varj \|^2_2 + \Lambda_{jj} )  - {\varj}^{\top} 
  \varnotj {(\varnotj^\top \varnotj + \lambdanotj)}^{-1} \varnotj^\top\varj \\
  \vperm &= -{ (\varnotj^\top\varnotj + \lambdanotj)}^{-1} \varnotj^\top \varj\\
  \betanotj &=  {(\varnotj^\top\varnotj + \lambdanotj)}^{-1} \varnotj^\top\by 
  \enspace.
\end{align*}
Indeed, using the Schur complement, one writes $\betaperm$
as follows
:
\begin{align*}
  \betaperm = \frac{1}{\aperm}
    \begin{pmatrix} 
      1 \\
     \vperm
    \end{pmatrix}
    \begin{pmatrix} 
    1 & {\vperm}^{\top}
    \end{pmatrix} \begin{pmatrix}
  {\varj}^{\top} \bfy \\
  \varnotj^{\top} \bfy
  \end{pmatrix} + 
  \begin{pmatrix} 
  0 \\
  \betanotj
  \end{pmatrix}
  \enspace.
\end{align*}
Hence, $\betaperm$ can be obtained as a correction of the vector of 
coefficients $\betanotj$ obtained under the smaller model. 
The key observation to be made here is that 
$\varj$ does not intervene
in the expression 
${(\varnotj^\top\varnotj + \lambdanotj)}^{-1}$ , 
which is the bottleneck in the computation of $\aperm$, $\vperm$ and
$\betanotj$. It can therefore be performed once for all permutations. 
Additionally, ${(\varnotj^\top\varnotj + \lambdanotj)}^{-1}$ can be
cheaply computed from $\left({\bfX}^{\top}\bfX + \bfLambda\right)^{-1}$ as
follows:
%
\begin{align*}
	{(\varnotj^\top\varnotj + \lambdanotj)}^{-1} = &
  \left[ \left({\bfX}^{\top}\bfX + \bfLambda\right)^{-1} \right]_{\fgebackslash j \fgebackslash j} - \\
  &
  \left[ \left({\bfX}^{\top}\bfX + \bfLambda\right)^{-1} \right]_{\fgebackslash j j}
  \left[ \left({\bfX}^{\top}\bfX + \bfLambda\right)^{-1} \right]_{j j}^{-1}
  \left[ \left({\bfX}^{\top}\bfX + \bfLambda\right)^{-1} \right]_{j{\fgebackslash j}}
  \enspace.
\end{align*}
Thus we compute $\left({\bfX}^{\top}\bfX + \bfLambda\right)^{-1}$ once,
firstly correct for the removal of variable $j$, secondly correct for 
permutation $b$, thus eventually requiring $O(B(q^{3}+q^{2}n)))$ operations.

\end{document}